\documentclass[12pt]{article}

\usepackage{graphicx,psfrag,epsf,color}
\usepackage{epsfig, graphicx}
\usepackage{hyperref}
\usepackage{multirow}
\usepackage{booktabs}
\usepackage{amsthm,amsmath,amssymb}
\usepackage{natbib}
\usepackage[linesnumbered,ruled,lined]{algorithm2e}
\usepackage{tikz}
\usepackage[figuresright]{rotating}
\setcitestyle{authoryear, round}
\theoremstyle{plain}

\newtheorem{assumption}{Assumption}
\newtheorem{theorem}{Theorem}
\newtheorem{corollary}{Corollary}
	
\newtheorem{remark}{Remark}	
\newtheorem{proposition}{Proposition}	

\usepackage{tikz}
\tikzstyle{every node}=[font=\large,scale=1.2]
\hypersetup{
	colorlinks=true,
	linkcolor=red,
	urlcolor=blue,
	citecolor=blue
}
\usepackage{graphicx}
\usepackage{booktabs}
\usepackage{diagbox} 
\usepackage{multirow}
\usepackage{makecell} 
\usepackage{longtable} 
\usepackage{array}
\usepackage{tabularx}
\usepackage{threeparttable}
\usepackage{xspace}

\author{%
  Tao Shen\\
  National University of Singapore\\
  Yifan Cui\thanks{Correspondence to Yifan Cui $<$\href{mailto:cuiyf@zju.edu.cn}{cuiyf@zju.edu.cn}$>$}\\
  Zhejiang University}
\title{Optimal Treatment Regimes for Proximal \\ Causal Learning}
\date{}

\begin{document}
	\maketitle

\begin{abstract}
  A common concern when a policymaker draws causal inferences from and makes decisions based on observational data is that the measured covariates are insufficiently rich to account for all sources of confounding, i.e., the standard no confoundedness assumption fails to hold.
The recently proposed proximal causal inference framework shows that proxy variables that abound in real-life scenarios can be leveraged to identify causal effects and therefore facilitate decision-making.
Building upon this line of work, we propose a novel optimal individualized treatment regime based on so-called outcome and treatment confounding bridges.  
We then show that the value function of this new optimal treatment regime is superior to that of existing ones in the literature.  
Theoretical guarantees, including identification, superiority, excess value bound, and consistency of the estimated regime, are established.  
Furthermore, we demonstrate the proposed optimal regime via numerical experiments and a real data application. 
\end{abstract}

\section{Introduction}

Data-driven individualized decision-making has received tremendous attention nowadays due to its applications in healthcare, economics, marketing, etc. A large branch of work has focused on maximizing the expected utility of implementing the estimated optimal policy over a target population based on randomized controlled trials or observational studies, e.g., \citet{athey2021policy,chakraborty2013statistical,jiang2019entropy,kitagawa2018should,kosorok2019precision,murphy2003optimal,qian2011performance,robins1986new,robins1994correcting,robins1997causal,tsiatis2019dynamic,wu2019matched,zhao2012estimating,zhao2019efficient}.

A critical assumption commonly made in these studies, known as unconfoundedness or exchangeability, precludes the existence of unmeasured confounding.
Relying on an assumed ability of the decision-maker to accurately measure covariates relevant to a variety of confounding mechanisms present in a given observational study, 
causal effects, value functions, and other relevant quantities can be nonparametrically identified. However, such an assumption might not always be realistic in observational studies or randomized trials subject to non-compliance \citep{robins1994correcting,robins1997causal}. Therefore, it is of great interest in recovering confounding mechanisms from measured covariates to infer causal effects and facilitate decision-making. 
A prevailing strand of work has been devoted to using instrumental variable \citep{angrist1996identification,imbens1994} as a proxy variable in dynamic treatment regimes and reinforcement learning settings \citep{Cui2021Individualized,cui2021semiparametric,cui2021necessary,han2019optimal,liao2021instrumental,pz,qiu2021optimal,stensrud2022optimal}.

Recently, Tchetgen Tchetgen et al. proposed the so-called proximal causal inference framework, a formal potential outcome framework for proximal causal learning, which while explicitly acknowledging covariate measurements as imperfect proxies of confounding mechanisms, establishes causal identification in settings where exchangeability on the basis of measured covariates fails. 
Rather than as current practice dictates, assuming that adjusting for all measured covariates, unconfoundedness can be attained, proximal causal inference essentially requires that the investigator can correctly classify a subset of measured covariates $L\in \mathcal L$ into three types: i) variables $X\in \mathcal X$ that may be common causes of the treatment and outcome variables; ii) treatment-inducing confounding proxies $Z\in \mathcal Z$; and iii) outcome-inducing confounding proxies $W\in \mathcal W$.

There is a fast-growing literature on proximal causal inference since it has been proposed
(\citealp{cui2020semiparametric,dukes2021proximal,ghassami2023partial,kompa2022deep,li2022double,mastouri2021proximal,miao2018confounding,shi2020selective,shi2021theory,shpitser2021proximal,singh2020kernel,tchetgen2020introduction,ying2021proximal,ying2022proximal} and many others). 
In particular,  \citet{miao2018identifying,tchetgen2020introduction} propose identification of causal effects through an outcome confounding bridge and \citet{cui2020semiparametric} propose identification 
through a treatment confounding bridge. 
A doubly robust estimation strategy \citep{chernozhukov2018double,robins1994estimation,rotnitzky1998semiparametric,scharfstein1999adjusting} is further proposed in 
\citet{cui2020semiparametric}.
In addition,  \citet{ghassami2022minimax} and \citet{kallus2021causal} propose a nonparametric estimation of causal effects through a min-max approach.
Moreover,  by adopting the proximal causal inference framework, \citet{qi2021proximal} consider optimal individualized treatment regimes (ITRs) estimation, \citet{sverdrup2023proximal} consider learning heterogeneous treatment effects, and \citet{bennett2021proximal} consider off-policy evaluation in partially observed Markov decision processes.

In this paper, we aim to estimate optimal ITRs under the framework of proximal causal inference. We start with reviewing two in-class ITRs that map from $\mathcal X \times \mathcal W$ to $\mathcal A$ and $\mathcal X \times \mathcal Z$ to $\mathcal A$, respectively, where $\mathcal A$ denotes the binary treatment space. The identification of value function and the learning strategy for these two optimal in-class ITRs are proposed in \citet{qi2021proximal}.  
In addition, \citet{qi2021proximal} also consider a maximum proximal learning optimal ITR that maps from $\mathcal{X}\times \mathcal{W} \times \mathcal{Z}$ to $\mathcal A$ with the ITRs being restricted to either $\mathcal{X}\times \mathcal{W} \rightarrow \mathcal A$ or $\mathcal{X}\times \mathcal{Z}\rightarrow \mathcal A$.
In contrast to their maximum proximal learning ITR, in this paper, we propose a brand new policy class whose ITRs map 
from measured covariates $\mathcal{X}\times \mathcal{W} \times \mathcal{Z}$ to $\mathcal A$, which incorporates the predilection between these two in-class ITRs. Identification and superiority of the proposed optimal ITRs compared to existing ones are further established. 

The main contributions of our work are four-fold. 
Firstly, by leveraging treatment and outcome confounding bridges under the recently proposed proximal causal inference framework, identification results regarding the proposed class $\mathcal{D}_{\mathcal{ZW}}^{\Pi}$ of ITRs that map $\mathcal{X}\times \mathcal{W} \times \mathcal{Z}$ to $\mathcal{A}$ are established. The proposed ITR class can be viewed as a generalization of existing ITR classes proposed in the literature. 
Secondly, an optimal subclass of $\mathcal{D}_{\mathcal{ZW}}^{\Pi}$ is further introduced. Learning optimal treatment regimes within this subclass leads to a superior value function.
Thirdly, we propose a learning approach to estimating the proposed optimal ITR. Our learning pipeline begins with the estimation of confounding bridges adopting the deep neural network method proposed by \citet{kompa2022deep}. Then we use optimal treatment regimes proposed in \citet{qi2021proximal} as preliminary regimes to estimate our optimal ITR.  
Lastly, we establish an excess value bound for the value difference between the estimated treatment regime and existing ones in the literature, and the consistency of the estimated regime is also demonstrated.


\section{Methodology}
\label{section2}
\subsection{Optimal individualized treatment regimes}
\label{riuc}
We briefly introduce some conventional notation for learning optimal ITRs.
Suppose $A$ is a binary variable representing a treatment option that takes values in the treatment space $\mathcal{A} = \{-1,1\}$. Let $L \in \mathcal{L}$ be a vector of observed covariates,  and $Y$ be the outcome of interest. Let $Y (1)$ and $Y (-1)$ be the potential outcomes under an intervention that sets the treatment to values $1$ and $-1$, respectively. Without loss of generality, we assume that larger values of $Y$ are preferred.

Suppose the following standard causal assumptions hold:  (1) Consistency: $Y = Y(A)$. That is, the observed outcome matches the potential outcome under the realized treatment. (2) Positivity: $\mathbb{P}(A = a|L) > 0$ for $a \in \mathcal{A}$ almost surely, i.e., both treatments are possible to be assigned. 

We consider an ITR class $\mathcal{D}$ containing ITRs that are measurable functions mapping from the covariate space $\mathcal{L}$ onto the treatment space $\mathcal A$. 
For any $d \in \mathcal{D}$, the potential outcome under a hypothetical intervention that assigns treatment according to $d$ is defined as 
\begin{equation*}
    Y(d(L)) \overset{\triangle}{=} Y(1)\mathbb{I}\{d(L) = 1\} + Y(-1)\mathbb{I}\{d(L) = -1\},
\end{equation*}
where $\mathbb{I}\{\cdot\}$ denotes the indicator function.  
The value function of ITR $d$ is defined as the expectation of the potential outcome, i.e.,
\begin{equation*}
    V(d) \overset{\triangle}{=}  \mathbb{E}[Y(d(L))].
\end{equation*}
It can be easily seen that an optimal ITR can be expressed as 
\begin{align*}
d^*(L) = \text{sign}\{ \mathbb{E}(Y(1)-Y(-1)|L)\}
\end{align*}
or
\begin{align*}
d^* = \arg\max_{d \in \mathcal D} \mathbb{E}[Y(d(L))].
\end{align*}

There are many ways to identify optimal ITRs under different sets of assumptions.
The most commonly seen assumption is the unconfoundedness: $Y(a) \perp A|L$ for $a=\pm 1$, i.e., upon conditioning on $L$, there is no unmeasured confounder affecting both $A$ and $Y$. 
Under this unconfoundedness assumption, the value function of a given regime $d$ can be identified by \citep{qian2011performance}
\begin{equation*}
    V(d) = \mathbb{E}\left[ \frac{Y\mathbb{I}\{A=d(L)\}}{f(A|L)} \right],
\end{equation*}
where $f(A|L)$ denotes the propensity score \citep{rosenbaum1983central}, and the optimal
ITR is identified by 
\begin{equation*}
    d^* = \arg\max_{d \in \mathcal{D}} V(d) = \arg \max_{d \in \mathcal{D}} \mathbb{E}\left[ \frac{Y\mathbb{I}\{A=d(L)\}}{f(A|L)} \right].
\end{equation*}
We refer to \citet{qian2011performance,zhang2012robust,zhao2012estimating} for more details of learning optimal ITRs in this unconfounded setting.

Because confounding by unmeasured factors cannot generally be ruled out with certainty in observational studies or randomized experiments subject to non-compliance, skepticism about the unconfoundedness assumption in observational studies is often warranted. 
To estimate optimal ITRs subject to potential unmeasured confounding, \citet{cui2021semiparametric} propose instrumental variable approaches to learning optimal ITRs. Under certain instrumental variable assumptions, the optimal ITR can be identified by
\begin{equation*}
    \arg \max_{d \in \mathcal{D}} \mathbb{E}\left[\frac{MAY\mathbb{I}\{A = d(L)\}}{\{\mathbb{P}(A=1|M=1,L)-\mathbb{P}(A=1|M=-1,L)\}f(M|L)}\right],
\end{equation*}
where $M$ denotes a valid binary instrumental variable. 
Other works including \citet{Cui2021Individualized,cui2021necessary,han2019optimal,pz} consider a sign or partial identification of causal effects to estimate suboptimal ITRs using instrumental variables.

\subsection{Existing optimal ITRs for proximal causal inference}
\label{existing}
Another line of research in causal inference considers negative control variables as proxies to mitigate confounding bias \citep{kuroki2014measurement, miao2018identifying, shi2020multiply,tchetgen2014control}. Recently, a formal potential outcome framework, namely proximal causal inference, has been developed by \citet{tchetgen2020introduction}, which has attracted tremendous attention since proposed.

\begin{figure}[htbp]
\centering
\begin{tikzpicture}[state/.style={circle, draw, minimum size=0.8cm}]
    \def\Ax{2}
    \def\Ay{0}
    \def\offset{2}
    \def\Bx{\Ax+5}
    \def\By{\Ay}
    \node[state, shape=circle,draw=black,font=\scriptsize, right of=s3] (s2) at (\Ax,\Ay) {$X$};
    \node[state, shape=circle,draw=black,font=\scriptsize] (s3) at (\Ax-0.4*\offset,\Ay) {$Z$};
    \node[state, shape=circle,draw=black,font=\scriptsize, right of=s2] (s5) at (\Ax+1*\offset,\Ay) {$W$};
    \node[state, shape=circle, draw=black,fill=lightgray,font=\scriptsize, above of=s2] (s4) at (\Ax+1.2,\Ay+0.5) {$U$};
     \node[state, shape=circle, draw=black,font=\scriptsize, below left of=s2] (s6) at (\Ax+0.8,\Ay-0.5) {$A$};
       \node[state, shape=circle, draw=black,font=\scriptsize, below right of=s2] (s7) at (\Ax+1.7,\Ay-0.5) {$Y$};
    
    \draw [-latex] (s2) to [right]  (s5);
    \draw [-latex] (s2) to [left]  (s3);
    \draw [-latex] (s4) to [above left]  (s2);
    \draw [-latex] (s4) to [above right]  (s5);
    \draw [-latex] (s4) to [above left]  (s3);
    \draw [-latex] (s2) to [above left]  (s6);
    \draw [-latex] (s2) to [above right]  (s7);
    \draw [-latex] (s4) to [above left]  (s6);
    \draw [-latex] (s4) to [above right]  (s7);
    \draw [-latex] (s6) to [right]  (s7);
    \draw [-latex] (s3) to [above right]  (s6);
    \draw [-latex] (s5) to [above left]  (s7);

\end{tikzpicture}
\caption{A causal DAG under the proximal causal inference framework.}
\label{fig:1}
\end{figure}
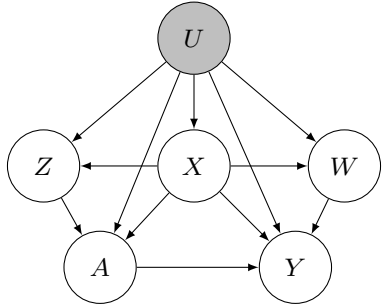

Following the proximal causal inference framework proposed in \citet{tchetgen2020introduction}, suppose that the measured covariate $L$ can be decomposed into three buckets $L = (X, W, Z)$, where $X \in \mathcal X$ affects both $A$ and $Y$, $W\in \mathcal W$ denotes an outcome-inducing confounding proxy that is a potential cause of the outcome which is related with the treatment
only through $(U, X)$, and $Z \in \mathcal Z$ is a treatment-inducing confounding proxy that 
is a potential cause of the treatment which is related with the outcome $Y$ through $(U, X, A)$. We now summarize several basic assumptions of the proximal causal inference framework.
\begin{assumption}
We make the following assumptions:\\
	(1) Consistency: $Y = Y(A,Z), \ W = W(A,Z)$.\\
	(2) Positivity: $\mathbb{P}(A = a\ |\ U, X) > 0, \ \forall a \in \mathcal{A}.$\\
	(3) Latent unconfoundedness: \\$(Z, A) \perp (Y (a), W )\ | \ U, X, \ \forall a \in \mathcal{A}.$
\label{std}
\end{assumption}
The consistency and positivity assumptions are conventional in the causal inference literature. The latent unconfoundedness essentially states that $Z$ cannot directly affect the outcome $Y$, and $W$ is not directly affected by either $A$ or $Z$. Figure~\ref{fig:1} depicts a classical setting that satisfies Assumption~\ref{std}. We refer to \citet{shi2020selective,tchetgen2020introduction} for other realistic settings for proximal causal inference. 

We first consider two in-class optimal ITRs that map from $\mathcal{X}\times \mathcal{Z}$ to $\mathcal{A}$ and $\mathcal{X}\times \mathcal{W}$ to $\mathcal{A}$, respectively. To identify optimal ITRs that map from $\mathcal{X}\times \mathcal{Z}$ to $\mathcal{A}$, we make the following assumptions. 
\begin{assumption}
	Completeness: For any $a \in \mathcal{A}, x \in \mathcal{X}$ and  square-integrable function $g$, $\mathbb{E}[g(U)\ | \ Z, A = a, X = x] = 0$ almost surely if and only if $g(U) = 0$ almost surely. 
\label{com1}
\end{assumption}
\begin{assumption}
  Existence of outcome confounding bridge: There exists an outcome confounding bridge function $h(w, a, x)$ that solves the following equation
  \begin{equation*}
  	\mathbb{E}[Y | Z, A, X] = \mathbb{E}[h(W,A,X) | Z, A, X],
  \end{equation*}
almost surely. 
\label{eofob}
\end{assumption}
The completeness Assumption \ref{com1} is a technical condition
central to the study of sufficiency in foundational theory of statistical inference. It essentially assumes that $Z$ has sufficient variability with respect to the variability of $U$.  We refer to \citet{tchetgen2020introduction} and \citet{miao2022identifying} for further discussions regarding the completeness condition. 
Assumption \ref{eofob} defines
a so-called inverse problem known as a Fredholm integral equation of the first kind through an outcome confounding bridge. 
The technical conditions for the existence of a solution to a Fredholm integral equation can be found in \citet{kress1989linear}.  

Let $\mathcal{D}_\mathcal{Z}$ be an ITR class that includes all measurable functions mapping from $\mathcal{X}\times \mathcal{Z}$ to $\mathcal{A}$.
 As shown in \citet{qi2021proximal}, under Assumptions \ref{std}, \ref{com1} and \ref{eofob}, for any $d_z \in \mathcal{D}_{\mathcal{Z}}$, the value function $V(d_z)$ can be nonparametrically identified by	
\begin{equation}
	V(d_z) = \mathbb{E}[h(W, d_z(X, Z), X)].
 \label{vd1}
\end{equation}
Furthermore, the in-class optimal treatment regime $d_z^* \in \arg\max_{d_z\in \mathcal{D}_{\mathcal{Z}}} V(d_z)$ is given by
\begin{equation*}
    d_z^*(X,Z) = \text{sign}\{\mathbb{E}[h(W,1,X) - h(W,-1,X)|X,Z]\}.
\end{equation*}

On the other hand, to identify optimal ITRs that map from $\mathcal{X}\times \mathcal{W}$ to $\mathcal{A}$, we make the following assumptions. 
\begin{assumption}
	Completeness: For any $a \in \mathcal{A}, x \in \mathcal{X}$ and square-integrable function $g$,  $\mathbb{E}[g(U)\ | \ W, A = a, X = x] = 0$  almost surely if and only if $g(U) = 0$   almost surely.
\label{com2}
\end{assumption}
\begin{assumption}
  Existence of treatment confounding bridge: There exists a treatment confounding bridge function $q(z, a, x)$ that solves the following equation
  \begin{equation*}
  	\frac{1}{\mathbb{P}(A = a | W, X)} = \mathbb{E}[q(Z, a, X) | W, A = a, X],
  \end{equation*}
   almost surely.
\label{eoftb}
\end{assumption}

Similar to Assumptions \ref{com1} and \ref{eofob}, 
Assumption \ref{com2} assumes that $W$ has sufficient variability relative to the variability of $U$, and Assumption \ref{eoftb} defines another Fredholm integral equation of the first kind through a treatment confounding bridge $q$. 

Let $\mathcal{D}_\mathcal{W}$ be another ITR class that includes all measurable functions mapping from $\mathcal{X}\times \mathcal{W}$ to $\mathcal{A}$.
As shown in \citet{qi2021proximal}, under Assumptions \ref{std}, \ref{com2} and \ref{eoftb}, for any $d_w \in \mathcal{D}_\mathcal{W}$, the value function $V(d_w)$ can be nonparametrically identified by	
\begin{equation}
	V(d_w) = \mathbb{E}[Yq(Z, A, X)\mathbb{I}\{d_w(X,W) = A\}].
\label{vd2}
\end{equation}
The in-class optimal treatment regime $d_w^* \in \arg\max_{d_w\in \mathcal{D}_\mathcal{W}} V(d_w)$ is given by
\begin{align*}
    &d_w^*(X,W) = \text{sign}\{\mathbb{E}[Yq(Z,1, X)\mathbb{I}\{A = 1\}- Yq(Z,-1, X)\mathbb{I}\{A = -1\}|X,W]\}.
\end{align*}

Moreover, \citet{qi2021proximal} consider the ITR class $\mathcal{D}_\mathcal{Z} \cup \mathcal{D}_\mathcal{W}$ and propose a maximum proximal learning optimal regime based on this ITR class.
For any $d_{z \cup w} \in \mathcal{D}_\mathcal{Z} \cup \mathcal{D}_\mathcal{W}$, under Assumptions~\ref{std}-\ref{eoftb}, the value function $V(d_{z \cup w})$ for any $d_{z \cup w} \in \mathcal{D}_\mathcal{Z} \cup \mathcal{D}_\mathcal{W}$ can be identified by 
\begin{align}
   &V(d_{z \cup w}) = \mathbb{I}\{d_{z \cup w} \in \mathcal{D}_\mathcal{Z}\} \mathbb{E}[h(W, d_{z \cup w}(X, Z), X)] + \mathbb{I}\{d_{z \cup w} \in \mathcal{D}_\mathcal{W}\} \mathbb{E}[Yq(Z, A, X)\mathbb{I}\{d_{z \cup w}(X,W) = A\}].
\label{zcupw}
\end{align} 
The optimal ITR within this class is given by $d_{z \cup w}^* \in \arg \max_{d_{z \cup w} \in \mathcal{D}_\mathcal{Z} \cup \mathcal{D}_\mathcal{W}} V(d_{z \cup w})$, and they show that the corresponding optimal value function takes the maximum value between two optimal in-class ITRs, i.e.,
\begin{equation*}
     V(d_{z \cup w}^*) = \max\{ V(d_z^*), V(d_w^*)\}.
\end{equation*}

\subsection{Optimal decision-making based on two confounding bridges}\label{sec:main}
As discussed in the previous section, given that neither $\mathbb{E} [Y (a) | X, U]$  nor $\mathbb{E} [Y (a) | X, W, Z ]$ for any $a \in \mathcal{A}$ may be identifiable under the proximal causal inference setting, one might nevertheless consider ITRs mapping from 
 $\mathcal{X} \times \mathcal{W}$ to $\mathcal{A}$, from $\mathcal{X} \times \mathcal{Z}$ to $\mathcal{A}$, from $\mathcal{X} \times \mathcal{W} \times \mathcal{Z}$ to $\mathcal{A}$
 as well as from $\mathcal{X}$ to $\mathcal{A}$.
Intuitively, policy-makers might want to use as much information as they can to facilitate their decision-making.
Therefore, ITRs mapping from $\mathcal{X} \times \mathcal{W} \times \mathcal{Z}$ to $\mathcal{A}$ are of great interest if information regarding $(X,W,Z)$ is available.

As a result, a natural question arises: is there an ITR mapping from $\mathcal{X} \times \mathcal{W} \times \mathcal{Z}$ to $\mathcal{A}$ which dominates existing ITRs proposed in the literature? In this section, we answer this question by proposing a novel optimal ITR and showing its superiority in terms of global welfare.   

We first consider the following class of ITRs that map from $\mathcal{X} \times \mathcal{W} \times \mathcal{Z}$ to $\mathcal{A}$, 
\begin{align*}
\mathcal{D}_{\mathcal{ZW}}^{\Pi} \overset{\triangle}{=} \{d_{zw}^{\pi}:	d_{zw}^{\pi}(X,W,Z) = \pi(X)d_z(X,Z) + (1-\pi(X))d_w(X,W), d_z \in \mathcal{D}_{\mathcal{Z}}, d_w \in \mathcal{D}_{\mathcal{W}},\pi \in \Pi \},
\end{align*}
where $\Pi$ is the policy class containing all measurable functions $\pi: \mathcal{X} \rightarrow \{0,1\}$ that indicate the individualized predilection between $d_z$ and $d_w$.

\begin{remark}
Note that $\mathcal{D}_{\mathcal{Z}}, \mathcal{D}_{\mathcal{W}}$ and $\mathcal{D}_{\mathcal{Z}}\cup\mathcal{D}_{\mathcal{W}}$ are subsets of $\mathcal{D}_{\mathcal{ZW}}^{\Pi}$  with a particular choice of $\pi$. For example, 
$\mathcal{D}_{\mathcal{Z}}$ is $\mathcal{D}_{\mathcal{ZW}}^{\Pi}$ with restriction on $\pi(X)= 1$;
$\mathcal{D}_{\mathcal{W}}$ is $\mathcal{D}_{\mathcal{ZW}}^{\Pi}$ with restriction on $\pi(X)= 0$; 
$\mathcal{D}_{\mathcal{Z}}\cup\mathcal{D}_{\mathcal{W}}$ is $\mathcal{D}_{\mathcal{ZW}}^{\Pi}$ with restriction on $\pi(X)=1$ or $\pi(X)=0$. 

\end{remark}

In the following theorem, we demonstrate that by leveraging the treatment and outcome confounding bridge functions, we can nonparametrically identify the value function over the policy class  $\mathcal{D}_{\mathcal{ZW}}^{\Pi}$, i.e., $V(d_{zw}^{\pi})$ for $d_{zw}^{\pi} \in \mathcal{D}_{\mathcal{ZW}}^{\Pi}$.

\begin{theorem}\label{th1}
 Under Assumptions~\ref{std}-\ref{eoftb}, for any $d_{zw}^{\pi} \in \mathcal{D}_{\mathcal{ZW}}^{\Pi}$, the value function $V(d_{zw}^{\pi})$ can be nonparametrically identified by
\begin{align}\label{eq:V}
	V(d_{zw}^{\pi}) = \mathbb{E}[\pi(X)h(W,d_z(X,Z),X) + (1-\pi(X))Yq(Z,A,X)\mathbb{I}\{d_w(X,W)=A\}].
\end{align}
\end{theorem}

One of the key ingredients of our constructed new policy class $\mathcal{D}_{\mathcal{ZW}}^{\Pi}$ is the choice of $\pi(\cdot)$. It suggests an individualized strategy for treatment decisions between the two given treatment regimes. 
Because we are interested in policy learning, a suitable choice of $\pi(\cdot)$ that leads to a larger value function is more desirable.
Therefore, we construct the following $\bar{\pi}(X;d_z,d_w)$, 
\begin{align}
	\bar{\pi}(X;d_z,d_w) \overset{\triangle}{=} \mathbb{I}\{\mathbb{E}[h(W,d_z(X,Z),X)|X] \geq \mathbb{E}[Yq(Z,A,X)\mathbb{I}\{d_w(X,W)=A\}|X]\}.
\label{indi}
\end{align}
In addition, given any $d_z \in \mathcal{D}_{\mathcal{Z}}$ and $d_w \in \mathcal{D}_{\mathcal{W}}$, we define 
\begin{align*}
d_{zw}^{\bar{\pi}} (X,W,Z) \overset{\triangle}{=}  \bar{\pi}(X;d_z,d_w) d_z(X,Z) +  (1-\bar{\pi}(X;d_z,d_w))d_w(X,W).
\end{align*}

We then obtain the following result which justifies the superiority of $\bar{\pi}$.
\begin{theorem}
\label{comd51}
Under Assumptions \ref{std}-\ref{eoftb}, for any $d_z \in \mathcal{D}_{\mathcal{Z}}$ and $d_w \in \mathcal{D}_{\mathcal{W}}$, 
\begin{align*}
		V(d_{zw}^{\bar{\pi}}) \geq \max\{V(d_z), V(d_w)\}.
\end{align*}
\end{theorem}

Theorem~\ref{comd51} establishes that for the particular choice of $\bar \pi$ given in~\eqref{indi}, the value function of $d_{zw}^{\bar{\pi}}$ is no smaller than that of 
$d_z$ and $d_w$ for any $d_z \in \mathcal D_{\mathcal Z}$, and $d_w \in \mathcal D_{\mathcal W}$. 
Consequently, Theorem~\ref{comd51} holds for $d_z^*$ and $d_w^*$. Hence, we propose the following optimal ITR $d_{zw}^{\bar{\pi}*}$,
\begin{align*}
    d_{zw}^{\bar{\pi}*}(X,W,Z) \overset{\triangle}{=}\bar{\pi}(X;d_z^*,d_w^*)d_z^*(X,Z) + (1-\bar{\pi}(X;d_z^*,d_w^*))d_w^*(X,W),
\end{align*}
and we have the following corollary.
\begin{corollary}
\label{comop}
Under Assumptions \ref{std}-\ref{eoftb}, we have that
     \begin{equation*}
		V(d_{zw}^{\bar{\pi}*}) \geq \max\{V(d_z^*), V(d_w^*), V(d_{z \cup w}^*)\}.
	\end{equation*} 
\end{corollary}

Corollary~\ref{comop} essentially states that the value of $d_{zw}^{\bar{\pi}*}$ dominates that of $d_z^*$, $d_w^*$, as well as $d_{z \cup w}^*$. 
Moreover, the proposition below demonstrates the optimality of $d_{zw}^{\bar{\pi}*}$ within the proposed class.
\begin{proposition}
\label{opti}
Under Assumptions \ref{std}-\ref{eoftb}, we have that
    $$d_{zw}^{\bar{\pi}*} \in \arg \max_{d_{zw}^{\pi} \in \mathcal{D}_{\mathcal{ZW}}^{\Pi}} V(d_{zw}^{\pi}).$$
\end{proposition}
Therefore, $d_{zw}^{\bar{\pi}*}$ is an optimal ITR of policymakers' interest.

\section{Statistical Learning and Optimization}
\label{section3}

\subsection{Estimation of the optimal ITR $d_{zw}^{\bar{\pi}*}$}
\label{dzdwl}
The estimation of $d_{zw}^{\bar{\pi}*}$ consists of four steps: (i) estimation of confounding bridges $h$ and $q$; (ii) estimation of preliminary ITRs $d_{z}^*$ and $d_w^*$; (iii) estimation of $\bar{\pi}(X;d_z^*,d_w^*)$; and (iv) learning $d_{zw}^{\bar{\pi}*}$ based on (ii) and (iii). 
The estimation problem (i) has been developed by \citet{cui2020semiparametric,miao2018confounding} using the generalized method of moments, \citet{ghassami2022minimax, kallus2021causal} by a min-max estimation \citep{dikkala2020minimax} using kernels, and \citet{kompa2022deep} using deep learning; and (ii) has been developed by \citet{qi2021proximal}. We restate estimation of (i) and (ii) for completeness.  
With regard to (i), recall that Assumptions \ref{eofob} and \ref{eoftb} imply the following conditional moment restrictions 
\begin{align*}
    \mathbb{E}[Y - h(W,A,X) | Z, A, X] =& 0,\\
    \mathbb{E}\left[ 1- \mathbb{I}\{A = a\}q(Z,a,X)|W, X\right] =& 0, \forall a \in \mathcal{A}. 
\end{align*}
respectively. \citet{kompa2022deep} propose a deep neural network approach to estimating bridge functions which avoids the reliance on kernel methods. We adopt this approach in our simulation and details can be found in the Appendix. 

To estimate $d_z^*$, we consider classification-based approaches according to \citet{zhang2012robust, zhao2012estimating}. Under Assumptions \ref{std}, \ref{com1} and \ref{eofob}, maximizing the value function in \eqref{vd1} is equivalent to minimizing the following classification error
 \begin{align}\label{eq:min}
    \mathbb{E}[\{h(W,1,X)-h(W,-1,X)\}\mathbb{I}\{d_z(X,Z) \neq 1\}]
\end{align}
over $d_z \in \mathcal{D}_\mathcal{Z}$.  By choosing some measurable decision function $g_z\in\mathcal{G}_\mathcal{Z} : \mathcal{X} \times \mathcal{Z} \rightarrow \mathbb{R}$, we let $d_z(X,Z) = \text{sign}(g_z(X,Z))$. We consider the following empirical version of \eqref{eq:min},
\begin{align*}
    \min_{g_z \in \mathcal{G}_z} \mathbb{P}_n[\{\hat{h}(W,1,X)-\hat{h}(W,-1,X)\} \mathbb{I}\{g_z(X,Z)<0\}].
\end{align*}
Due to the non-convexity and non-smoothness of the sign operator, we replace the sign operator with a smooth surrogate function and adopt the hinge loss $\phi(x) = \max\{1 - x, 0\}$.  By adding a penalty term  $\rho_{z}||g_z||_{\mathcal{G}_\mathcal{Z}}^2$ to avoid overfitting, we solve
\begin{align}\label{eq:minz}
    \hat{g}_z &\in \arg\min_{g_z \in \mathcal{G}_z} \mathbb{P}_n[\{\hat{h}(W,1,X)-\hat{h}(W,-1,X)\} \phi(g_z(X,Z))] + \rho_{z}||g_z||_{\mathcal{G}_\mathcal{Z}}^2,
\end{align}
where $\rho_z >0$ is a tuning parameter.
 The estimated ITR then follows $\hat{d}_z(X,Z) = \text{sign}(\hat{g}_z(X,Z))$. Similarly, under Assumptions~\ref{std}, \ref{com2} and \ref{eoftb}, maximizing the value function in \eqref{vd2} is equivalent to minimizing the following classification error
\begin{align*}
    \mathbb{E}[\{Yq(Z,1,X)\mathbb{I}\{A=1\}-Yq(Z,-1,X)\mathbb{I}\{A=-1\}\}\mathbb{I}\{d_w(X,W) \neq 1\}]
\end{align*}
over $d_w \in \mathcal{D}_\mathcal{W}$. By the same token, the problem is transformed into minimizing the following empirical error
\begin{align}\label{eq:minw}
     \hat{g}_w &\in \arg\min_{g_w \in \mathcal{G}_\mathcal{W}} \mathbb{P}_n[\{Y\hat{q}(Z,1,X)\mathbb{I}\{A=1\} -Y\hat{q}(Z,-1,X)\mathbb{I}\{A=-1\}\} \phi(g_w(X,W))] + \rho_{w}||g_w||_{\mathcal{G}_\mathcal{W}}^2.
\end{align}
The estimated ITR is obtained via $\hat{d}_w(X,W) = \text{sign}(\hat{g}_w(X,W))$. 

For problem (iii), given two preliminary ITRs, we construct an estimator $\hat{\pi}(X;\hat{d}_z,\hat{d}_w)$, that is, for $x \in \mathcal{X}$,
\begin{align}
\hat{\pi}(x;\hat{d}_z,\hat{d}_w) = \mathbb{I}\{\hat{\delta}(x;\hat{d}_z,\hat{d}_w)\geq 0 \}, \nonumber
\end{align}
where $\hat{\delta}(x;\hat{d}_z,\hat{d}_w)$ denotes a generic estimator of \begin{align*}
\delta(x;\hat{d}_z,\hat{d}_w) \overset{\triangle}{=} \mathbb{E}[h(W,\hat{d}_z(X,Z),X) - Yq(Z,A,X)\mathbb{I}\{\hat{d}_w(X,W)=A\}|X = x],
\end{align*}
where the expectation is taken with respect to everything except $\hat{d}_z$ and $\hat{d}_w$. 
For example, the Nadaraya-Watson kernel regression estimator \citep{nadaraya1964estimating} can be used, i.e., $\hat{\delta}(x;\hat{d}_z,\hat{d}_w)$ is expressed as 
\begin{align*}
\frac{\sum_{i=1}^n \{\hat{h}(W_i, \hat{d}_z(x, Z_i),x)-Y_i\hat{q}(Z_i,A_i,x)\mathbb{I}\{\hat{d}_w(x,W_i)=A_i\}\}K(\frac{||x - X_i||_2}{\gamma})}{\sum_{i=1}^n K(\frac{||x - X_i||_2}{\gamma})},
\end{align*}
where $K: \mathbb{R} \rightarrow \mathbb{R}$ is a kernel function such as Gaussian kernel, $||\cdot||_2$ denotes the $L_2$-norm, and $\gamma$ denotes the bandwidth.

Finally, given $\hat{d}_z, \hat{d}_w$ and $\hat{\pi}(X;\hat{d}_z,\hat{d}_w)$, $\hat{d}_{zw}^{\hat{\pi}}$ is estimated by the following plug-in regime,
\begin{align}\label{eq:est}
\hat{d}_{zw}^{\hat{\pi}}(X,W,Z) = \hat{\pi}(X;\hat{d}_z,\hat{d}_w)	\hat{d}_z(X,Z) +(1-\hat{\pi}(X;\hat{d}_z,\hat{d}_w))\hat{d}_w(X,W). 
\end{align}

\subsection{Theoretical guarantees for $\hat{d}^{\hat{\pi}}_{zw}$}
\label{sec:theory}
In this subsection, we first present an optimality guarantee for the estimated ITR $\hat{d}^{\hat{\pi}}_{zw}$ in terms of its value function 
\begin{align*}
  V(\hat{d}_{zw}^{\hat{\pi}}) = \mathbb{E}[\hat{\pi}(X;\hat{d}_z,\hat{d}_w) h(W, \hat{d}_z(X,Z),X) + (1- \hat{\pi}(X;\hat{d}_z,\hat{d}_w)) Y q(Z,A,X)\mathbb{I}\{\hat{d}_w(X,W) = A\}],
\end{align*}
where the expectation is taken with respect to everything except $\hat{\pi}$, $\hat{d}_z$ and $\hat{d}_w$.

We define an oracle optimal ITR which assumes $\bar{\pi}(X;\hat{d}_z,\hat{d}_w)$ is known, 
\begin{align*}
\hat{d}_{zw}^{\bar{\pi}}(X,W,Z) \overset{\triangle}{=} \bar{\pi}(X;\hat{d}_z,\hat{d}_w) \hat{d}_z(X,Z) + (1-\bar{\pi}(X;\hat{d}_z,\hat{d}_w)) \hat{d}_w(X,W).
\end{align*}
The corresponding value function of this oracle optimal ITR is given by
\begin{align*}
  V(\hat{d}_{zw}^{\bar{\pi}}) = \mathbb{E}[\bar{\pi}(X;\hat{d}_z,\hat{d}_w) h(W, \hat{d}_z(X,Z),X) + (1- \bar{\pi}(X;\hat{d}_z,\hat{d}_w)) Y q(Z,A,X)\mathbb{I}\{\hat{d}_w(X,W) = A\}],
\end{align*}
where the expectation is taken with respect to everything except $\hat{d}_z$ and $\hat{d}_w$.

Then the approximation error incurred by estimating $\hat{\pi}(X;\hat{d}_z,\hat{d}_w)$ is given by
\begin{align*}
    &\mathbb{K}(\hat{\pi}) \overset{\triangle}{=}  V(\hat{d}_{zw}^{\bar{\pi}}) 
 - V(\hat{d}_{zw}^{\hat{\pi}}).
\end{align*}
Moreover, we define the following gain
\begin{align*}
 \mathbb{G}(\bar{\pi}) \overset{\triangle}{=} \min\{ V(\hat{d}_{zw}^{\bar{\pi}}) - V(\hat{d}_z), V(\hat{d}_{zw}^{\bar{\pi}}) - V(\hat{d}_w)\}.
\end{align*}
It is clear that this gain $\mathbb{G}(\bar{\pi})$ by introducing $\bar{\pi}$ is always non-negative as indicated by Theorem \ref{comd51}.
Then we have the following excess value bound for the value of $\hat{d}^{\hat{\pi}}_{zw}$ compared to existing ones in the literature.
\begin{proposition}
\label{hate}
Under Assumptions \ref{std}-\ref{eoftb},
    \begin{align*}
		V(\hat{d}^{\hat{\pi}}_{zw}) =  \max\{V(\hat{d}_z), V(\hat{d}_w)\} - \mathbb{K}(\hat{\pi}) + \mathbb{G}(\bar{\pi})= V(\hat{d}_{z\cup w}) - \mathbb{K}(\hat{\pi}) + \mathbb{G}(\bar{\pi}).
	\end{align*}
\end{proposition}

Proposition~\ref{hate} establishes a link between the value function of the estimated ITR $\hat{d}^{\hat{\pi}}_{zw}$, and that of $\hat{d}_z$, $\hat{d}_w$, and $\hat{d}_{z\cup w}$.
As shown in Appendix~\ref{sec:asy}, $\mathbb{K}(\hat{\pi})$ diminishes as the sample size increases, therefore, $\hat{d}^{\hat{\pi}}_{zw}$ has a significant improvement compared to other optimal ITRs depending on the magnitude of $ \mathbb{G}(\bar{\pi})$. 

Furthermore, we establish the consistency of the proposed regime based on the following assumption, which holds for example when $\hat{d}_z$ and $\hat{d}_w$
are estimated using indirect methods.
\begin{assumption}
    For $\hat{d}_z, \hat{d}_w$, $E[h(W,\hat{d}_z(X,Z),X)|X]-E[h(W,d_z^*(X,Z),X)|X]=o_p(n^{-\xi})$ almost surely and $ E[Yq(Z,A,X)\mathbb{I}\{\hat{d}_w(X,W) = A\}|X]-E[Yq(Z,A,X)\mathbb{I}\{d_w^*(X,W) = A\}|X]=o_p(n^{-\varphi})$ almost surely.
\label{aszw}
\end{assumption}
\begin{proposition}
\label{consistency}
    Under Assumptions \ref{std}-\ref{aszw}, we have $V(\hat{d}_{zw}^{\hat{\pi}}) \xrightarrow{p} V(d_{zw}^{\bar{\pi}*})$.
\end{proposition}

\section{Numerical Experiments}
\label{section5}

The data generating mechanism for $(X,A,Z,W,U)$ follows the setup proposed in \citet{cui2020semiparametric} and is summarized in Appendix~\ref{sec:dgp}. 
To evaluate the performance of the proposed framework, we vary $b_1(X)$, $b_2(X)$, $b_3(X)$, $b_a$  and $b_w$ in $\mathbb{E}[Y|X,A,Z,W,U]$ to incorporate heterogeneous treatment effects including the settings considered in \citet{qi2021proximal}. 
The adopted data generating mechanism is compatible with the following $h$ and $q$,
$$h(W,A,X)=b_0 +\left\{b_1(X)+ b_aW + b_3(X)W\right\}\frac{1+A}{2} +b_wW + b_2(X)X,$$
$$q(Z,A,X) = 1+ \exp\left\{At_0+At_zZ +t_a\frac{1+A}{2} + At_xX\right\},$$
where $t_0 = 0.25, t_z = -0.5, t_a = -0.125$, and $t_x = (0.25, 0.25)^T$.
We derive preliminary optimal ITRs $d_{z}^*$ and $d_{w}^*$ in Appendix~\ref{sec:dstar}, from which we can see that $X,Z,W$ are relevant variables for individualized decision-making.

We consider six scenarios in total, and the setups of varying parameters are deferred to Appendix~\ref{sec:dgp}. 
For each scenario, training datasets $\{Y_i, A_i, X_i, Z_i, W_i\}_{i=1}^n$ are generated following the above mechanism with a sample size $n = 1000$. For each training dataset, we then apply the aforementioned methods to learn the optimal ITR.
In particular, the preliminary ITRs $\hat{d}_z$ and $\hat{d}_w$ are estimated using a linear decision rule, and $\hat{\pi}(x;\hat{d}_z,\hat{d}_w)$ is estimated using a Gaussian kernel. More details can be found in the Appendix~\ref{sec:im}.

To evaluate the estimated treatment regimes, we consider the following generating mechanism for testing datasets: $X \sim \mathcal{N}(\Gamma_x, \Sigma_x)$, 
\begin{align*}
(Z, W, U)|X &\sim \mathcal{N} \Bigg\{ \left ( \begin{array}{clr}
\alpha_0+\alpha_a p_a+\alpha_x X\\
\mu_0+\mu_a p_a+\mu_x X\\
\kappa_0+\kappa_a p_a+\kappa_x X\\
\end{array}\right), 
&\Sigma = \left( \begin{array}{clr}
\sigma^2_z & \sigma_{zw} & \sigma_{zu}\\
\sigma_{zw} & \sigma_{w}^2 & \sigma_{wu}\\
\sigma_{zu} & \sigma_{wu} & \sigma_{u}^2
\end{array}
\right) \Bigg\},
\end{align*}
where the parameter settings can be found in Appendix~\ref{sec:dgp}. The testing dataset is generated with a size $10000$, and the empirical value function for the estimated ITR is used as a performance measure. The simulations are replicated 200 times. 
To validate our approach and demonstrate its superiority, we have also computed empirical values for other optimal policies, including existing optimal ITRs for proximal causal inference, as discussed in Section~ \ref{existing}, along with optimal ITRs generated through causal forest \citep{athey2019estimating} and outcome weighted learning \citep{zhao2012estimating}.

\begin{figure*}[ht]
        \centering
        \includegraphics[width=\textwidth, height = 0.6\textwidth]{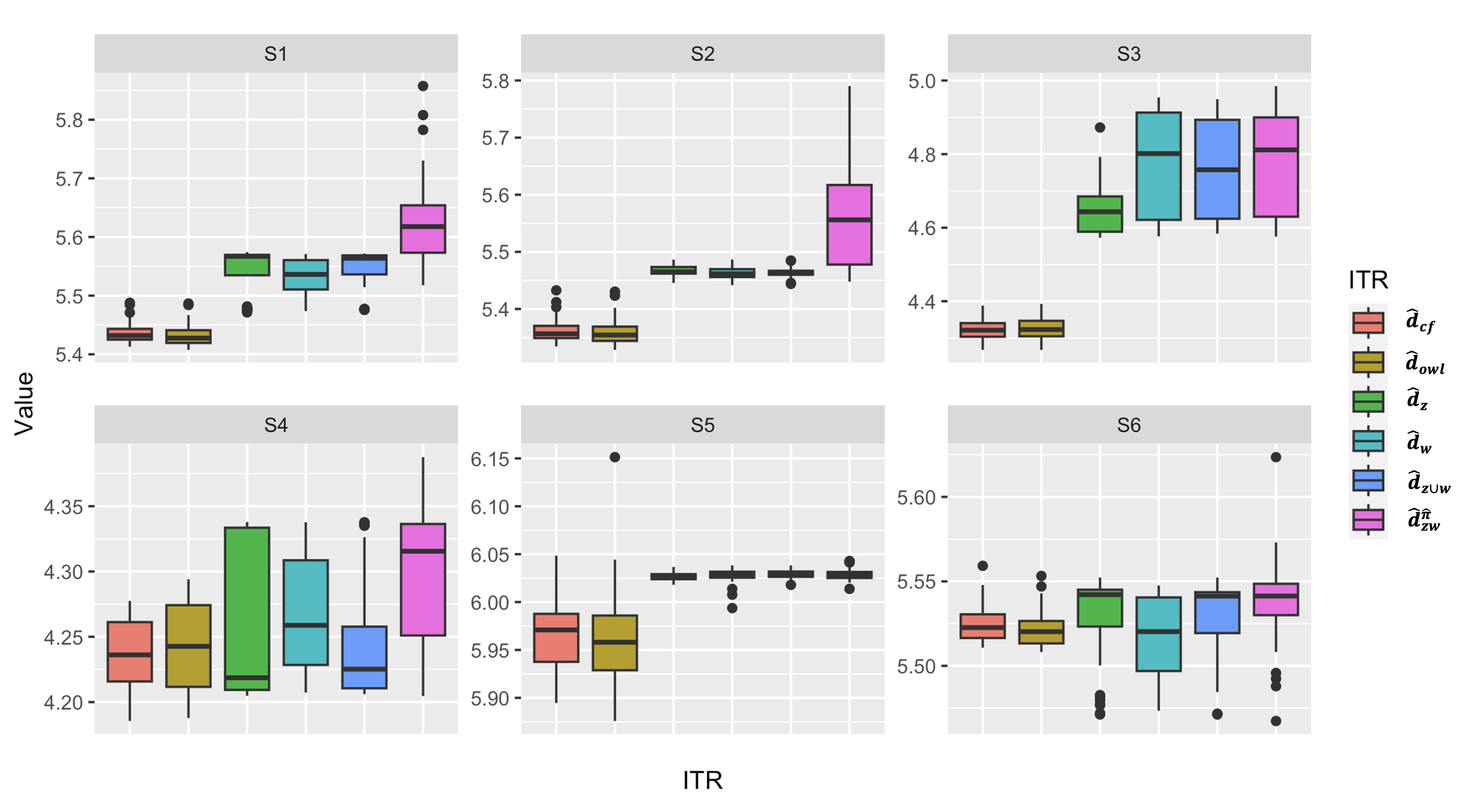}
        \caption{Boxplots of the empirical value functions ($\hat{d}_{cf}$ and $\hat{d}_{owl}$ denote estimated ITRs using causal forest and outcome weighted learning respectively).}
        \label{fig}
\end{figure*}

Figure~\ref{fig} presents the empirical value functions of different optimal ITRs for the six scenarios. As expected, $\hat{d}_z, \hat{d}_w, \hat{d}_{z\cup w},$ and $\hat{d}_{zw}^{\hat{\pi}}$ consistently outperform $\hat{d}_{cf}$ and $\hat{d}_{owl}$, which highlights their effectiveness in addressing unmeasured confounding. Meanwhile, across all scenarios, $\hat{d}_{zw}^{\hat{\pi}}$ yields superior or comparable performance compared to the other estimated treatment regimes, which justifies the statements made in Sections \ref{section2} and \ref{section3}. In addition, as can be seen in Scenario 5, 
all ITRs relying on the proximal causal inference framework perform similarly, which is not surprising as $\hat{d}_z(X,Z)$ and $\hat{d}_w(X,W)$ agree for most subjects. To further underscore the robust performance of our proposed approach, we include additional results with a changed sample size and a modified behavior policy in Appendix~\ref{sec:add}.

\section{Real Data Application}\label{sec:real}
In this section, we demonstrate the proposed optimal ITR via a real dataset originally designed to measure the effectiveness of right heart catheterization 
(RHC) for ill patients in intensive care units (ICU), under the Study to Understand Prognoses and Preferences for Outcomes and Risks of Treatments (SUPPORT, \citet{connors1996effectiveness}). These data have been re-analyzed in a number of papers in both causal inference and survival analysis literature with assuming unconfoundednss \citep{cui2019selective,tan2006distributional,tan2020model,tan2019regularized,2015biasreduce} or accounting for unmeasured confounding \citep{cui2020semiparametric,lin1998assessing,qi2021proximal,tchetgen2020introduction,ying2022proximal}.

There are 5735 subjects included in the dataset, in which 2184 were treated (with $A=1$) and 3551 were untreated (with $A = -1$). The outcome $Y$ is the duration from admission to death or censoring.  Overall, 3817 patients survived and 1918 died within 30
days. Following \citet{tchetgen2020introduction}, we collect 71 covariates including demographic factors, diagnostic information, estimated survival probability, comorbidity, vital signs, physiological status, and functional status (see \citet{hirano2001estimation} for additional discussion on covariates). Confounding in this study stems from the fact that ten physiological status measures obtained from blood tests conducted at the initial phase of admission may be susceptible to significant measurement errors. Furthermore, besides the lab measurement errors, whether other unmeasured confounding factors exist is unknown to the data analyst. Because variables measured from these tests offer only a single snapshot of the underlying physiological condition, they have the potential to act as confounding proxies. We consider a total of four settings, varying the number of selected proxies from 4 to 10. Within each setting, treatment-inducing proxies are first selected based on their strength of association with the treatment (determined through logistic regression of $A$ on $L$), and outcome-inducing proxies are then chosen based on their association with the outcome (determined through linear regression of $Y$ on $A$ and $L$). Excluding the selected proxy variables, other measured covariates are included in $X$. We then estimate $\hat{d}_z, \hat{d}_w, \hat{d}_{z \cup w}$, and $\hat{d}_{zw}^{\hat{\pi}}$ using the SUPPORT dataset in a manner similar to that described in Section \ref{section5}, with the goal optimizing the patients' 30-day survival after their entrance into the ICU. 

\begin{figure*}[ht]
        \centering
        \includegraphics[width=\textwidth, height = 0.12\textwidth]{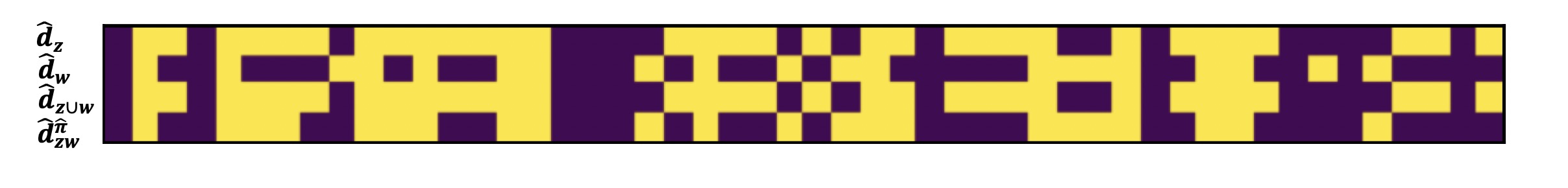}
        \caption{Graphical representation of concordance between estimated ITRs.}
        \label{fig3}
\end{figure*}

The estimated value functions of our proposed ITR, alongside existing ones, are summarized in Appendix~\ref{sec:add2}. As can be seen, our proposed regime has the largest value among all settings. For a visual representation of the concordance between the estimated optimal ITRs, we refer to Figure~\ref{fig3} (results from Setting 1). The horizontal ordinate represents the 50 selected subjects and the vertical axis denotes the decisions made from corresponding ITRs. 
The purple and yellow blocks stand for being recommended treatment values of -1 and 1 respectively.  
For the subjects with purple or yellow columns, $\hat{d}_z(X,Z) = \hat{d}_w(X,W)$, which leads to the same treatment decision for the other two ITRs. 
For columns with mixed colors, $\hat{d}_z(X,Z)$ and $\hat{d}_w(X,W)$ disagree.
We see that in this case $\hat{d}_{z\cup w}(X,W,Z)$ always agree with $\hat{d}_z(X,Z)$, while $\hat{d}_{zw}^{\hat{\pi}}(X,W,Z)$ take values from $\hat{d}_z(X,Z)$ or $\hat{d}_w(X,W)$ depending on the individual criteria of the subjects as indicated by $\hat{\pi}$. In addition to the quantitative analysis, we have also conducted a qualitative assessment of the estimated regime to validate its performance. For further details, please refer to Appendix~\ref{sec:add2}.

\section{Discussion}
\label{section6}
We acknowledge several limitations of our work. Firstly, the proximal causal inference framework relies on the validity of treatment- and outcome-inducing confounding proxies. When the assumptions are violated, the proximal causal inference estimators can be biased even if unconfoundedness on the basis of measured covariates in fact holds. Therefore, one needs to carefully sort out proxies especially when domain knowledge is lacking. Secondly, while the proposed regime significantly improves upon existing methods both theoretically and numerically, it is not yet shown to be the sharpest under our considered model. It is still an open question to figure out if a more general policy class could be considered. Thirdly, our established theory provides consistency and superiority of our estimated regime. It is of great interest to derive convergence rates for Propositions \ref{hate} and \ref{consistency} following \cite{jiang2017uniform}. In addition, it may be challenging to develop inference results for the value function of the estimated optimal treatment regimes, and further studies are warranted. 

\section*{Acknowledgement}
Yifan Cui was supported by the National Natural Science Foundation of China.


\newpage
\bibliographystyle{abbrvnat}
\bibliography{main}
\newpage
\appendix
\begin{center}
{\Large \textbf{Supplementary Material}}
\end{center}
\section{Proof of identification \eqref{zcupw}}

The proof is straightforward. We state it here for clarity and completeness.
Note that
$$\mathbb{I}\{d_{z\cup w}(X,W,Z)=1\} = \mathbb{I}\{d_{z\cup w} \in \mathcal{D}_\mathcal{Z}\}\mathbb{I}\{d_{z\cup w}(X,Z)=1\} + \mathbb{I}\{d_{z\cup w} \in \mathcal{D}_\mathcal{W}\}\mathbb{I}\{d_{z\cup w}(X,W)=1\},$$
$$\mathbb{I}\{d_{z\cup w}(X,W,Z)=-1\} = \mathbb{I}\{d_{z\cup w} \in \mathcal{D}_\mathcal{Z}\}\mathbb{I}\{d_{z\cup w}(X,Z)=-1\} + \mathbb{I}\{d_{z\cup w} \in \mathcal{D}_\mathcal{W}\}\mathbb{I}\{d_{z\cup w}(X,W)=-1\}.$$
Therefore, we have 
\begin{align*}
\mathbb{E}[Y(1)\mathbb{I}\{d_{z\cup w}(X,W,Z)=1\}]	&= \mathbb{E}[Y(1)\mathbb{I}\{d_{z\cup w} \in \mathcal{D}_\mathcal{Z}\}\mathbb{I}\{d_{z\cup w}(X,Z)=1\} \\
&+ Y(1)\mathbb{I}\{d_{z\cup w} \in \mathcal{D}_\mathcal{W}\}\mathbb{I}\{d_{z\cup w}(X,W)=1\}]\\
&= \mathbb{I}\{d_{z\cup w} \in \mathcal{D}_\mathcal{Z}\} \mathbb{E}[Y(1)\mathbb{I}\{d_{z\cup w}(X,Z)=1\}] \\
&+\mathbb{I}\{d_{z\cup w} \in \mathcal{D}_\mathcal{W}\}\ \mathbb{E}[Y(1)\mathbb{I}\{d_{z\cup w}(X,W)=1\}].
\end{align*}
Similarly, 
\begin{align*}
\mathbb{E}[Y(-1)\mathbb{I}\{d_{z\cup w}(X,W,Z)=-1\}]	 
&= \mathbb{I}\{d_{z\cup w} \in \mathcal{D}_\mathcal{Z}\} \mathbb{E}[Y(-1)\mathbb{I}\{d_{z\cup w}(X,Z)=-1\}] \\
&+\mathbb{I}\{d_{z\cup w} \in \mathcal{D}_\mathcal{W}\}\ \mathbb{E}[Y(-1)\mathbb{I}\{d_{z\cup w}(X,W)=-1\}].
\end{align*}
So
\begin{align*}
V(d_{z \cup w }) &= \mathbb{E}[Y(1)\mathbb{I}\{d_{z\cup w}(X,W,Z)=1\}] + 	\mathbb{E}[Y(-1)\mathbb{I}\{d_{z\cup w}(X,W,Z)=-1\}]	 \\
&= \mathbb{I}\{d_{z\cup w} \in \mathcal{D}_\mathcal{Z}\}  \mathbb{E}[Y(1)\mathbb{I}\{d_{z\cup w}(X,Z)=1\} + Y(-1)\mathbb{I}\{d_{z\cup w}(X,Z)=-1\}]	\\
&+ \mathbb{I}\{d_{z\cup w} \in \mathcal{D}_\mathcal{W}\}  \mathbb{E}[Y(1)\mathbb{I}\{d_{z\cup w}(X,W)=1\} + Y(-1)\mathbb{I}\{d_{z\cup w}(X,W)=-1\}]	\\
&= \mathbb{I}\{d_{z\cup w} \in \mathcal{D}_\mathcal{Z}\}  \mathbb{E}[h(W,d_{z\cup w}(X,Z),X)] +\mathbb{I}\{d_{z\cup w} \in \mathcal{D}_\mathcal{W}\}  \mathbb{E}[Yq(Z,A,X)\mathbb{I}\{d_{z\cup w}(X,W)=A\}],
\end{align*}
where the last equality holds due to identification results \eqref{vd1} and \eqref{vd2}.

\section{Proof of Theorem \ref{th1}}

Recall that 
$V(d_{zw}^{\pi}) = \mathbb{E}[Y(1)\mathbb{I}\{d_{zw}^{\pi}(X,W,Z)=1\}+Y(-1)\mathbb{I}\{d_{zw}^{\pi}(X,W,Z)=-1\}]$, we essentially need to consider the first term $\mathbb{E}[Y(1)\mathbb{I}\{d_{zw}^{\pi}(X,W,Z)=1\}]$. Note that
$$\mathbb{I}\{d_{zw}^{\pi}(X,W,Z)=1\} = \mathbb{I}\{\pi(X) = 1\}\mathbb{I}\{d_z(X,Z) = 1\} + \mathbb{I}\{\pi(X) = 0\}\mathbb{I}\{d_w(X,W) = 1\},$$
we have 
\begin{align*}
\mathbb{E}[Y(1)\mathbb{I}\{d_{zw}^{\pi}(X,W,Z)=1\}]	&= \mathbb{E}[Y(1)\mathbb{I}\{\pi(X) = 1\}\mathbb{I}\{d_z(X,Z) = 1\} + Y(1)\mathbb{I}\{\pi(X) = 0\}\mathbb{I}\{d_w(X,W) = 1\}]\\
&= \mathbb{E}[\mathbb{I}\{\pi(X) = 1\}\mathbb{E}[Y(1)\mathbb{I}\{d_z(X,Z) = 1\}|X]\\
&+ \mathbb{I}\{\pi(X) = 0\}\mathbb{E}[Y(1)\mathbb{I}\{d_w(X,W) = 1\}|X]].
\end{align*}
By leveraging the outcome confounding bridge, we have
\begin{align*}
\mathbb{E}[Y(1)\mathbb{I}\{d_z(X,Z) = 1\}|X]	 
&= \mathbb{E}[\mathbb{E}[Y(1)|X,Z]\mathbb{I}\{d_z(X,Z) = 1\}|X]\\
&= \mathbb{E}[\mathbb{E}[\mathbb{E}[Y(1)|X,Z,U]|X,Z]\mathbb{I}\{d_z(X,Z) = 1\}|X]\\
&= \mathbb{E}[\mathbb{E}[\mathbb{E}[Y|X,U,A=1]|X,Z]\mathbb{I}\{d_z(X,Z) = 1\}|X]\\
&= \mathbb{E}[\mathbb{E}[\mathbb{E}[h(W,1,X)|X,U]|X,Z]\mathbb{I}\{d_z(X,Z) = 1\}|X]\\
&= \mathbb{E}[\mathbb{E}[\mathbb{E}[h(W,1,X)|X,Z,U]|X,Z]\mathbb{I}\{d_z(X,Z) = 1\}|X]\\
&= \mathbb{E}[h(W,1,X)\mathbb{I}\{d_z(X,Z) = 1\}|X],
\end{align*}
where the third equality is due to Assumption~\ref{std}, the fourth equality can be verified by Theorem 1 in \cite{miao2018identifying} under Assumptions~\ref{com1} and \ref{eofob}, and the fifth equality is due to Assumption~\ref{std}.
Moreover, by leveraging the treatment confounding bridge, we have
\begin{align*}
\mathbb{E}[Y(1)\mathbb{I}\{d_w(X,W) = 1\}|X]	 
&= \mathbb{E}[\mathbb{E}[Y(1)|X,W]\mathbb{I}\{d_w(X,W) = 1\}|X]\\
&= \mathbb{E}[\mathbb{E}[\mathbb{E}[Y(1)|X,W,U]|X,W]\mathbb{I}\{d_w(X,W) = 1\}|X]\\
&= \mathbb{E}[\mathbb{E}[\mathbb{E}[Y(1)|X,W,U,A=1]|X,W]\mathbb{I}\{d_w(X,W) = 1\}|X]\\
&= \mathbb{E}[\mathbb{E}[\mathbb{E}[Y(1)|X,W,U,A=1]\mathbb{E}[q(Z,1,X)|X,U,A=1] \\
&\ \ \  \mathbb{P}(A=1|X,U)|X,W] \mathbb{I}\{d_w(X,W) = 1\}|X]\\
&= \mathbb{E}[\mathbb{E}[\mathbb{E}[Yq(Z,1,X)\mathbb{I}\{A=1\}|X,U,W]|X,W]\mathbb{I}\{d_w(X,W) = 1\}|X]\\
&= \mathbb{E}[Yq(Z,1,X)\mathbb{I}\{A=1\}\mathbb{I}\{d_w(X,W) = 1\}|X],
\end{align*}
 where the third equality is due to Assumption~\ref{std}, the fourth equality is implied by Theorem 2.2 of \cite{cui2020semiparametric}  under Assumptions~\ref{com2} and \ref{eoftb}, and the fifth equality is due to Assumption~\ref{std}. Therefore,
\begin{align}\label{pf1}
\mathbb{E}[Y(1)\mathbb{I}\{d_{zw}^{\pi}(X,W,Z)=1\}]	
&= \mathbb{E}[\mathbb{I}\{\pi(X) = 1\}\mathbb{E}[Y(1)\mathbb{I}\{d_z(X,Z) = 1\}|X] \nonumber \\
&+ \mathbb{I}\{\pi(X) = 0\}\mathbb{E}[Y(1)\mathbb{I}\{d_w(X,W) = 1\}|X]] \nonumber  \\
&= \mathbb{E}[\mathbb{I}\{\pi(X) = 1\}\mathbb{E}[h(W,1,X)\mathbb{I}\{d_z(X,Z) = 1\}|X]\nonumber \\
&+ \mathbb{I}\{\pi(X) = 0\}\mathbb{E}[Yq(Z,1,X)\mathbb{I}\{A=1\}\mathbb{I}\{d_w(X,W) = 1\}|X]]\nonumber  \\
&=  \mathbb{E}[\mathbb{I}\{\pi(X) = 1\}h(W,1,X)\mathbb{I}\{d_z(X,Z) = 1\} \nonumber \\
&+ \mathbb{I}\{\pi(X) = 0\}Yq(Z,1,X)\mathbb{I}\{A=1\}\mathbb{I}\{d_w(X,W) = 1\}].
\end{align}
Similarly, as 
$$\mathbb{I}\{d_{zw}^{\pi}(X,W,Z)=-1\} = \mathbb{I}\{\pi(X) = 1\}\mathbb{I}\{d_z(X,Z) = -1\} + \mathbb{I}\{\pi(X) = 0\}\mathbb{I}\{d_w(X,W) = -1\},$$
we have
\begin{align}\label{pf2}
& \mathbb{E}[Y(-1)\mathbb{I}\{d_{zw}^{\pi}(X,W,Z)=-1\}] \nonumber \\
&=  \mathbb{E}[\mathbb{I}\{\pi(X) = 1\}h(W,-1,X)\mathbb{I}\{d_z(X,Z) = -1\} \nonumber \\
&+ \mathbb{I}\{\pi(X) = 0\}Yq(Z,-1,X)\mathbb{I}\{A=-1\}\mathbb{I}\{d_w(X,W) = -1\}].
\end{align}
Combining \eqref{pf1} and \eqref{pf2}, we have
\begin{align*}
V(d_{zw}^{\pi}) &= \mathbb{E}[Y(1)\mathbb{I}\{d_{zw}^{\pi}(X,W,Z)=1\}+Y(-1)\mathbb{I}\{d_{zw}^{\pi}(X,W,Z)=-1\}]\\
&=	\mathbb{E}[\mathbb{I}\{\pi(X) = 1\}h(W,1,X)\mathbb{I}\{d_z(X,Z) = 1\} \\
&+ \mathbb{I}\{\pi(X) = 0\}Yq(Z,1,X)\mathbb{I}\{A=1\}\mathbb{I}\{d_w(X,W) = 1\}\\
&+\mathbb{I}\{\pi(X) = 1\}h(W,-1,X)\mathbb{I}\{d_z(X,Z) = -1\} \\
&+ \mathbb{I}\{\pi(X) = 0\}Yq(Z,-1,X)\mathbb{I}\{A=-1\}\mathbb{I}\{d_w(X,W) = -1\}]\\
&= \mathbb{E}[\mathbb{I}\{\pi(X)=1\}h(W,d_z(X,Z),X)+\mathbb{I}\{\pi(X) = 0\}Yq(Z,A,X)\mathbb{I}\{d_w(X,W) = A\}]\\
&= \mathbb{E}[\pi(X)h(W,d_z(X,Z),X)+(1-\pi(X))Yq(Z,A,X)\mathbb{I}\{d_w(X,W) = A\}],
\end{align*}
which completes the proof.

\section{Proof of Theorem \ref{comd51}}
\label{proof_th2}
For any $d_z \in \mathcal D_{\mathcal Z}$ and $d_w \in \mathcal D_{\mathcal W}$, we have
\begin{align*}
	V(d_{zw}^{\bar{\pi}}) &= \mathbb{E}[\bar{\pi}(X;d_z,d_w) h(W,d_z(X,Z),X) + (1-\bar{\pi}(X;d_z,d_w)) Yq(Z,A,X)\mathbb{I}\{d_w(X,W)=A\}]\\
	&= \mathbb{E}[\bar{\pi}(X;d_z,d_w)\mathbb{E}[h(W,d_z(X,Z),X)|X] + (1-\bar{\pi}(X;d_z,d_w)) \mathbb{E}[Yq(Z,A,X)\mathbb{I}\{d_w(X,W)=A\}|X]]\\
	&= \mathbb{E}[\max\{\mathbb{E}[h(W,d_z(X,Z),X)|X], \mathbb{E}[Yq(Z,A,X)\mathbb{I}\{d_w(X,W)=A\}|X]\}],
\end{align*}
where the last equality is due to the definition of $\bar{\pi}(X;d_z,d_w)$. As
$$\max\{\mathbb{E}[h(W,d_z(X,Z),X)|X], \mathbb{E}[Yq(Z,A,X)\mathbb{I}\{d_w(X,W)=A\}|X]\} \geq \mathbb{E}[h(W,d_z(X,Z),X)|X],$$
and
\begin{align*}
\max\{\mathbb{E}[h(W,d_z(X,Z),X)|X], \mathbb{E}[Yq(Z,A,X)\mathbb{I}\{d_w(X,W)=A\}|X]\} \geq \mathbb{E}[Yq(Z,A,X)\mathbb{I}\{d_w(X,W)=A\}|X],
\end{align*}
taking expectations on both sides, we have
\begin{align*}
	&\mathbb{E}[\max\{\mathbb{E}[h(W,d_z(X,Z),X)|X], \mathbb{E}[Yq(Z,A,X)\mathbb{I}\{d_w(X,W)=A\}|X]\}]  \\
    &\geq \mathbb{E}[\mathbb{E}[h(W,d_z(X,Z),X)|X]]\\
	&= \mathbb{E}[h(W,d_z(X,Z),X)] \\
	&= V(d_z),\\
	& \mathbb{E}[\max\{\mathbb{E}[h(W,d_z(X,Z),X)|X], \mathbb{E}[Yq(Z,A,X)\mathbb{I}\{d_w(X,W)=A\}|X]\}]  \\
  &\geq \mathbb{E}[\mathbb{E}[Yq(Z,A,X)\mathbb{I}\{d_w(X,W)=A\}|X]]\\
	&= \mathbb{E}[Yq(Z,A,X)\mathbb{I}\{d_w(X,W)=A\}] \\
	&= V(d_w).
\end{align*}
Therefore, we have $V(d_{zw}^{\bar{\pi}}) \geq \max\{V(d_z), V(d_w)\}$.

\section{Proof of Corollary \ref{comop}}

Recall that 
$$d_{zw}^{\bar{\pi}*}(X,W,Z)  = \bar{\pi}(X;d_z^*, d_w^*) d_z^*(X,Z) + (1-\bar{\pi}(X;d_z^*, d_w^*))d_w^*(X,W),$$ 
with 
$$\bar{\pi}(X;d_z^*, d_w^*) = \mathbb{I}\{\mathbb{E}[h(W,d_z^*(X,Z),X|X] \geq \mathbb{E}[Yq(Z,A,X)\mathbb{I}\{d_w^*(X,W)=A\}|X]\},$$
we apply the result in Theorem \ref{comd51}, i.e.,
$$V(d_{zw}^{\bar{\pi}*}) \geq \max\{V(d_z^*), V(d_w^*)\}. $$
Due to $V(d_{z \cup w}^*) =  \max\{V(d_z^*), V(d_w^*)\}$ as shown in \cite{qi2021proximal}, we then conclude that $V(d_{zw}^{\bar{\pi}*}) \geq \max\{V(d_z^*), V(d_w^*), V(d_{z \cup w}^*)\}.$

\section{Proof of Proposition \ref{opti}}\label{sec:optclass}
In the following, we show  
$$d_{zw}^{\bar{\pi}*} \in \arg \max_{d_{zw}^{\pi} \in \mathcal{D}_{\mathcal{ZW}}^{\Pi}} V(d_{zw}^{\pi}).$$
Recall that 
\begin{align*}
V(d_{zw}^{\bar{\pi}}) &= \mathbb{E}[\bar{\pi}(X;d_z,d_w) h(W,d_z(X,Z),X) + (1-\bar{\pi}(X;d_z,d_w)) Yq(Z,A,X)\mathbb{I}\{d_w(X,W)=A\}]\\
	&= \mathbb{E}[\bar{\pi}(X;d_z,d_w)\mathbb{E}[h(W,d_z(X,Z),X)|X] + (1-\bar{\pi}(X;d_z,d_w)) \mathbb{E}[Yq(Z,A,X)\mathbb{I}\{d_w(X,W)=A\}|X]]\\
	&= \mathbb{E}[\max\{\mathbb{E}[h(W,d_z(X,Z),X)|X], \mathbb{E}[Yq(Z,A,X)\mathbb{I}\{d_w(X,W)=A\}|X]\}]\\
	&\geq \mathbb{E}[\pi(X)\mathbb{E}[h(W,d_z(X,Z),X)|X] + (1-\pi(X)) \mathbb{E}[Yq(Z,A,X)\mathbb{I}\{d_w(X,W)=A\}|X]]\\
	&= V(d_{zw}^{\pi}),
\end{align*}
for any $d_z \in \mathcal{D}_{\mathcal Z}, d_w \in \mathcal{D}_{\mathcal W}$ and $\pi(\cdot)$.
Therefore, we essentially need to show
$$d_{zw}^{\bar{\pi}*} \in \arg \max_{d_{zw}^{\bar{\pi}} \in \mathcal{D}^{\bar{\pi}}_{\mathcal{ZW}}} V(d_{zw}^{\bar{\pi}}),$$
where $\mathcal{D}^{\bar{\pi}}_{\mathcal{ZW}}\overset{\triangle}{=} \{d_{zw}^{\bar{\pi}}:	d_{zw}^{\bar{\pi}}(X,W,Z) = \bar{\pi}(X)d_z(X,Z) + (1-\bar{\pi}(X))d_w(X,W), d_z \in \mathcal{D}_{\mathcal{Z}}, d_w \in \mathcal{D}_{\mathcal{W}} \}$.
Recall that 
\begin{align*}
d_z^*(X,Z) = & \text{sign}\{\mathbb{E}[h(W,1,X)-h(W,-1,X)|X,Z]\},\\
d_w^*(X,W) = & \text{sign}\{\mathbb{E}[Yq(Z,A,X)\mathbb{I}\{A=1\}-Yq(Z,A,X)\mathbb{I}\{A=-1\}|X,W]\},
\end{align*}
we have 
\begin{align*}
\mathbb{E}[h(W,d_z^*(X,Z),X)|X,Z] \geq & \mathbb{E}[h(W,d_z(X,Z),X)|X,Z],\\
\mathbb{E}[Yq(Z,A,X)\mathbb{I}\{A=d_w^*(X,W)\}|X,W] \geq & \mathbb{E}[Yq(Z,A,X)\mathbb{I}\{A=d_w(X,W)\}|X,W].
\end{align*}
Taking expectation with respect to $Z$ and $W$ given $X$ respectively, we have
\begin{align}
\mathbb{E}[\mathbb{E}[h(W,d_z^*(X,Z),X)|X,Z]|X] \geq & \mathbb{E}[\mathbb{E}[h(W,d_z(X,Z),X)|X,Z]|X],
\label{op1}\\
\mathbb{E}[\mathbb{E}[Yq(Z,A,X)\mathbb{I}\{d_w^*(X,W)=A\}|X,W]|X] \geq & \mathbb{E}[\mathbb{E}[Yq(Z,A,X)\mathbb{I}\{d_w(X,W)=A\}|X,W]|X].
\label{op2}
\end{align}

By the proof given in Section~\ref{proof_th2}, we have that
\begin{align*}
	V(d_{zw}^{\bar{\pi}*}) &= \mathbb{E}[\max\{\mathbb{E}[h(W,d_z^*(X,Z),X)|X], \mathbb{E}[Yq(Z,A,X)\mathbb{I}\{d_w^*(X,W)=A\}|X]\}], 
\end{align*}
$$V(d_{zw}^{\bar{\pi}}) = \mathbb{E}[\max\{\mathbb{E}[h(W,d_z(X,Z),X)|X], \mathbb{E}[Yq(Z,A,X)\mathbb{I}\{d_w(X,W)=A\}|X]\}].$$
Therefore,
$$ V(d_{zw}^{\bar{\pi}*}) =\mathbb{E}[\max\{\mathbb{E}[\mathbb{E}[h(W,d_z^*(X,Z),X)|X,Z]|X],  \mathbb{E}[\mathbb{E}[Yq(Z,A,X)\mathbb{I}\{d_w^*(X,W)=A\}|X,W]|X]\}],$$
$$ V(d_{zw}^{\bar{\pi}}) = \mathbb{E}[\max\{\mathbb{E}[\mathbb{E}[h(W,d_z(X,Z),X)|X,Z]|X],  \mathbb{E}[\mathbb{E}[Yq(Z,A,X)\mathbb{I}\{d_w(X,W)=A\}|X,W]|X]\}].$$
From~\eqref{op1} and \eqref{op2}, we have $V(d_{zw}^{\bar{\pi}*}) \geq V(d_{zw}^{\bar{\pi}})$ for any $d_{zw}^{\bar{\pi}} \in \mathcal{D}^{\bar{\pi}}_{\mathcal{ZW}}$, which implies that $d_{zw}^{\bar{\pi}*}$ is the maximizer of $V(d_{zw}^{\pi})$.

\section{Proof of Proposition \ref{hate}}

By definition of $\mathbb{K}(\hat{\pi})$ we have
\begin{align}
V(\hat{d}^{\hat{\pi}}_{zw}) = V(\hat{d}_{zw}^{\bar{\pi}}) - \mathbb{K}(\hat{\pi}).
\label{th31}
\end{align}
Then from
\begin{align*}
 \mathbb{G}(\bar{\pi}) = \min\{ V(\hat{d}_{zw}^{\bar{\pi}}) - V(\hat{d}_z), V(\hat{d}_{zw}^{\bar{\pi}}) - V(\hat{d}_w)\},
\end{align*}
we can see
\begin{align}
\max\{V(\hat{d}_z),V(\hat{d}_w)\} = V(\hat{d}_{zw}^{\bar{\pi}}) -  \mathbb{G}(\bar{\pi}).
\label{th32}
\end{align}
Finally, combining \eqref{th31} and \eqref{th32}, we have
\begin{align*}
V(\hat{d}^{\hat{\pi}}_{zw}) &= \max\{V(\hat{d}_z), V(\hat{d}_w)\}  - \mathbb{K}(\hat{\pi}) +  \mathbb{G}(\bar{\pi}).
\end{align*}
As $V(\hat{d}_{z\cup w}) = \max\{V(\hat{d}_z), V(\hat{d}_w)\}$, we conclude that
\begin{align*}
V(\hat{d}^{\hat{\pi}}_{zw}) &= \max\{V(\hat{d}_z), V(\hat{d}_w)\}  - \mathbb{K}(\hat{\pi}) +  \mathbb{G}(\bar{\pi}) = V(\hat{d}_{z\cup w}) - \mathbb{K}(\hat{\pi}) +  \mathbb{G}(\bar{\pi}).
\end{align*}

\section{Asymptotics of $\mathbb{K}(\hat{\pi})$}\label{sec:asy}

Throughout this section, we assume that $X \in [0, 1]^p$ 
has a bounded density $f(x)$ and $\max\{|Y|, ||h||_{\infty}, ||q||_{\infty}\} \leq M$ for some $M>0$. In addition, we assume that 
$\sup_{w,a,x}|\hat h(w,a,x)- h(w,a,x)| = o_p(n^{-\alpha})$, $\sup_{z,a,x}|\hat q(z,a,x)- q(z,a,x)| = o_p(n^{-\beta})$ for some $\alpha,\beta >0$ \citep{chen2013optimal}. 
 Given the training dataset, we define an oracle estimator of $\delta(x;\hat{d}_z,\hat{d}_w)$
$$\delta'(x;\hat{d}_z,\hat{d}_w) \overset{\triangle}{=} \frac{\sum_{i=1}^n \{h(W_i, \hat{d}_z(x, Z_i),x)-Y_iq(Z_i,A_i,x)\mathbb{I}\{\hat{d}_w(x,W_i)=A_i\}\}K(\frac{||x - X_i||}{\gamma})}{\sum_{i=1}^n K(\frac{||x - X_i||}{\gamma})}.$$
We assume that with probability larger than $1-1/n$, for any $d_z\in \mathcal D_{\mathcal Z}$ and $d_w \in \mathcal D_{\mathcal W}$,
$\sup_x|\delta(x;d_z,d_w) - \delta'(x;d_z,d_w)| \leq  C_1 n^{-\gamma}$  for some $C_1>0$ and $\gamma>0$ under certain conditions \citep{jiang2017uniform}.
If we further impose a restriction on the carnality of preliminary policy classes and assume $|\mathcal D_{\mathcal Z}| = o(n)$ and $|\mathcal D_{\mathcal W}| = o(n)$, by a straightforward calculation, we have
$\sup_{x,d_z\in \mathcal D_{\mathcal Z},d_w\in \mathcal D_{\mathcal W}}|\hat \delta(x;d_z,d_w)-\delta(x;d_z,d_w)| \leq C_2 n^{-\zeta}$ on a set $\mathcal X_0$ and $\mathbb P(\mathcal X_0^c) \rightarrow 0$,
where $C_2>0, \zeta=\min\{\alpha,\beta,\gamma\}$, and $\mathcal X_0^c$ is the complement of $\mathcal X_0$.

To streamline the presentation, in the following, we abbreviate $\delta(X;\hat{d}_z,\hat{d}_w)$, $ \delta'(X;\hat{d}_z,\hat{d}_w)$ and $\hat{\delta}(X;\hat{d}_z,\hat{d}_w)$ as $\delta(X), \delta'(X)$ and $\hat{\delta}(X)$, respectively. 
Two subsets of $\mathcal{X}$, namely $\mathcal{X}_{f1}$ and $\mathcal{X}_{f2}$, are defined as
\begin{align*}
\mathcal{X}_{f1} &= \{x \in \mathcal{X}: \mathbb{I}\{\hat{\delta}(x) \geq 0\} = 1, \mathbb{I}\{\delta(x) \geq 0\} = 0\},\\
\mathcal{X}_{f2} &= \{x\in \mathcal{X}: \mathbb{I}\{\hat{\delta}(x) \geq 0\} = 0, \mathbb{I}\{\delta(x) \geq 0\} = 1\},
\end{align*} 
and we also define the complement set $\mathcal{X}_c$ as
\begin{align*}
\mathcal{X}_c &= \{x\in \mathcal{X}: \text{sign}(\hat{\delta}(x)) = \text{sign}(\delta(x))\},
\end{align*} 
with $\text{sign}(0) = 1$.
We see that $\mathcal{X}_{f1} \cap \mathcal{X}_{f2} = \emptyset,$ $\mathcal{X}_{f1} \cap \mathcal{X}_{c} = \emptyset$, $\mathcal{X}_{f2} \cap \mathcal{X}_{c} = \emptyset$,  and $\mathcal{X}_{f1} \cup \mathcal{X}_{f2} \cup \mathcal{X}_{c} = \mathcal{X}$.
From the definition of $\mathbb{K}(\hat{\pi})$, we have
\begin{align*}
    &\mathbb{K}(\hat{\pi}) = \int_{x \in \mathcal{X}} \mathbb{E}[\bar{\pi}(X;\hat{d}_z,\hat{d}_w) h(W, \hat{d}_z(X,Z),X) + (1- \bar{\pi}(X;\hat{d}_z,\hat{d}_w)) Y q(Z,A,X)\mathbb{I}\{\hat{d}_w(X,W) = A\}  \\  
    &-\hat{\pi}(X;\hat{d}_z,\hat{d}_w) h(W, \hat{d}_z(X,Z),X) + (1- \hat{\pi}(X;\hat{d}_z,\hat{d}_w)) Y q(Z,A,X)\mathbb{I}\{\hat{d}_w(X,W) = A\}|X = x] f(x)dx\\
    &=\int_{x \in \mathcal{X}_{f1}} -\delta(x) f(x)dx + \int_{x \in \mathcal{X}_{f2}} \delta(x) f(x)dx + \int_{x \in \mathcal{X}_{c}} 0 f(x)dx \\
    &= -\int_{x \in \mathcal{X}_{f1}} \delta(x) f(x)dx + \int_{x \in \mathcal{X}_{f2}} \delta(x) f(x)dx.
\end{align*}
The second equation holds because if $x \in \mathcal{X}_{f1}$, $\hat{\pi}(x;\hat{d}_z,\hat{d}_w) = 1, \bar{\pi}(x;\hat{d}_z,\hat{d}_w) = 0$ and
$$\mathbb{E}[\bar{\pi}(X;\hat{d}_z,\hat{d}_w) h(W, \hat{d}_z(X,Z),X) + (1- \bar{\pi}(X;\hat{d}_z,\hat{d}_w)) Y q(Z,A,X)\mathbb{I}\{\hat{d}_w(X,W) = A\} $$
$$- \hat{\pi}(X;\hat{d}_z,\hat{d}_w) h(W, \hat{d}_z(X,Z),X) + (1- \hat{\pi}(X;\hat{d}_z,\hat{d}_w)) Y q(Z,A,X)\mathbb{I}\{\hat{d}_w(X,W) = A\}|X = x] = -\delta(x);$$
if $x \in \mathcal{X}_{f2},$  $\hat{\pi}(x;\hat{d}_z,\hat{d}_w) = 0, \bar{\pi}(x_f;\hat{d}_z,\hat{d}_w) = 1$ and
$$\mathbb{E}[\bar{\pi}(X;\hat{d}_z,\hat{d}_w) h(W, \hat{d}_z(X,Z),X) + (1- \bar{\pi}(X;\hat{d}_z,\hat{d}_w)) Y q(Z,A,X)\mathbb{I}\{\hat{d}_w(X,W) = A\} $$
$$- \hat{\pi}(X;\hat{d}_z,\hat{d}_w) h(W, \hat{d}_z(X,Z),X) + (1- \hat{\pi}(X;\hat{d}_z,\hat{d}_w)) Y q(Z,A,X)\mathbb{I}\{\hat{d}_w(X,W) = A\}|X = x] = \delta(x);$$ 
if $x \in \mathcal{X}_c$, $\hat{\pi}(x;\hat{d}_z,\hat{d}_w) = \bar{\pi}(x_f;\hat{d}_z,\hat{d}_w)$ and
$$\mathbb{E}[\bar{\pi}(X;\hat{d}_z,\hat{d}_w) h(W, \hat{d}_z(X,Z),X) + (1- \bar{\pi}(X;\hat{d}_z,\hat{d}_w)) Y q(Z,A,X)\mathbb{I}\{\hat{d}_w(X,W) = A\} $$
$$- \hat{\pi}(X;\hat{d}_z,\hat{d}_w) h(W, \hat{d}_z(X,Z),X) + (1- \hat{\pi}(X;\hat{d}_z,\hat{d}_w)) Y q(Z,A,X)\mathbb{I}\{\hat{d}_w(X,W) = A\}|X = x] = 0.$$ 

Therefore, we essentially need to bound $-\int_{x \in \mathcal{X}_{f1}} \delta(x) f(x)dx$ and $\int_{x \in \mathcal{X}_{f2}} \delta(x) f(x)dx$ follows a similar proof. 
In this regard, we further split $\mathcal{X}_{f1}$ to $\mathcal{X}_{f1,1} = \{x \in \mathcal{X}_{f1}: \delta(x) \in (-C n^{-\zeta},C n^{-\zeta})\}$ and $\mathcal{X}_{f1,2}=\{x \in \mathcal{X}_{f1}: \delta(x) \notin (-C n^{-\zeta},C n^{-\zeta})\}$. Then it is easy to see that $-\int_{x \in \mathcal{X}_{f1,1}} \delta(x) f(x)dx$ is bounded by $O(n^{-\zeta})$ and 
$\mathbb{P}(\mathcal{X}_{f1,2})$ converges to 0 as $\mathbb P(\mathcal X_0^c)$ converges to 0.
We then conclude that 
$\mathbb{K}(\hat{\pi}) =o(1)$ almost surely.

\section{Proof of Proposition \ref{consistency}}
We start with defining two subsets of $\mathcal{X}$,
\begin{align*}
\mathcal{X}_{g1} &= \{x \in \mathcal{X}: \bar{\pi}(x,\hat{d}_z,\hat{d}_w) = 1, \bar{\pi}(x,d_z^*,d_w^*) = 0\},\\
\mathcal{X}_{g2} &= \{x \in \mathcal{X}: \bar{\pi}(x,\hat{d}_z,\hat{d}_w) = 0, \bar{\pi}(x,d_z^*,d_w^*) = 1\},
\end{align*} 
and we also define the complement set $\mathcal{X}_{gc}$ as
\begin{align*}
\mathcal{X}_{gc} &= \{x\in \mathcal{X}:\bar{\pi}(x,\hat{d}_z,\hat{d}_w) = \bar{\pi}(x,d_z^*,d_w^*) \},
\end{align*} 
which can also be split into
\begin{align*}
\mathcal{X}_{gc1} &= \{x\in \mathcal{X}:\bar{\pi}(x,\hat{d}_z,\hat{d}_w) = \bar{\pi}(x,d_z^*,d_w^*) = 0 \},\\
\mathcal{X}_{gc2} &= \{x\in \mathcal{X}:\bar{\pi}(x,\hat{d}_z,\hat{d}_w) = \bar{\pi}(x,d_z^*,d_w^*) = 1 \}.
\end{align*} 

We see that $\mathcal{X}_{g1} \cap \mathcal{X}_{g2} = \emptyset, \mathcal{X}_{gc1} \cap \mathcal{X}_{gc2} = \emptyset, \mathcal{X}_{gc1} \cup \mathcal{X}_{gc2} = \mathcal{X}_{gc}$, $\mathcal{X}_{g1} \cap \mathcal{X}_{gc} = \emptyset$, $\mathcal{X}_{g2} \cap \mathcal{X}_{gc} = \emptyset$,  and $\mathcal{X}_{g1} \cup \mathcal{X}_{g2} \cup \mathcal{X}_{gc} = \mathcal{X}$.

From the definition of $V(d_{zw}^{\bar{\pi}*})$ and $V(\hat{d}_{zw}^{\bar{\pi}})$, we have
\begin{align*}
    &V(d_{zw}^{\bar{\pi}*}) - V(\hat{d}_{zw}^{\bar{\pi}}) \\
    &= \int_{x \in \mathcal{X}} \mathbb{E}[\bar{\pi}(X;d_z^*,d_w^*) h(W, d_z^*(X,Z),X) + (1- \bar{\pi}(X;d_z^*,d_w^*)) Y q(Z,A,X)\mathbb{I}\{d_w^*(X,W) = A\}\\
    &- \bar{\pi}(X;\hat{d}_z,\hat{d}_w) h(W, \hat{d}_z(X,Z),X) - (1- \bar{\pi}(X;\hat{d}_z,\hat{d}_w)) Y q(Z,A,X)\mathbb{I}\{\hat{d}_w(X,W) = A\}|X = x] f(x)dx\\
    &=\int_{x \in \mathcal{X}_{g1}} E[Y q(Z,A,X)\mathbb{I}\{d_w^*(X,W) = A\} - h(W, \hat{d}_z(X,Z),X)|X=x] f(x)dx \\
    &+ \int_{x \in \mathcal{X}_{g2}} E[h(W, d_z^*(X,Z),X) - Y q(Z,A,X)\mathbb{I}\{\hat{d}_w(X,W) = A\}|X = x] f(x)dx \\
    &+ \int_{x \in \mathcal{X}_{gc1}}  E[Y q(Z,A,X)\mathbb{I}\{d_w^*(X,W) = A\} - Y q(Z,A,X)\mathbb{I}\{\hat{d}_w(X,W) = A\}|X = x]f(x)dx \\
    &+ \int_{x \in \mathcal{X}_{gc2}}  E[h(W, d_z^*(X,Z),X) - h(W, \hat{d}_z(X,Z),X)|X = x]f(x)dx.
\end{align*}

Then it is easy to see that 
$$\int_{x \in \mathcal{X}_{gc1}}  E[Y q(Z,A,X)\mathbb{I}\{d_w^*(X,W) = A\} - Y q(Z,A,X)\mathbb{I}\{\hat{d}_w(X,W) = A\}|X = x]f(x)dx$$
and 
$$\int_{x \in \mathcal{X}_{gc2}}  E[h(W, d_z^*(X,Z),X) - h(W, \hat{d}_z(X,Z),X)|X = x]f(x)dx$$
converge to 0 in probability according to Assumption~\ref{aszw}. 

Therefore, we essentially need to bound 
$$\int_{x \in \mathcal{X}_{g1}} E[Y q(Z,A,X)\mathbb{I}\{d_w^*(X,W) = A\} - h(W, \hat{d}_z(X,Z),X)|X=x] f(x)dx$$ and 
$$\int_{x \in \mathcal{X}_{g2}} E[h(W, d_z^*(X,Z),X) - Y q(Z,A,X)\mathbb{I}\{\hat{d}_w(X,W) = A\}|X = x] f(x)dx.$$ 
We further split $\mathcal{X}_{g1}$ to $$\mathcal{X}_{g1,1} = \{x \in \mathcal{X}_{g1}: E[Y q(Z,A,X)\mathbb{I}\{d_w^*(X,W) = A\} - h(W, \hat{d}_z(X,Z),X)|X=x] \in (-C n^{-\eta},C n^{-\eta})\}$$ and $$\mathcal{X}_{g1,2}=\{x \in \mathcal{X}_{g1}: E[Y q(Z,A,X)\mathbb{I}\{d_w^*(X,W) = A\} - h(W, \hat{d}_z(X,Z),X)|X=x] \notin (-C n^{-\eta},C n^{-\eta})\}$$ where $\eta = \min\{\xi, \varphi\}$. Then it is easy to see that $$\int_{x \in \mathcal{X}_{g1,1}} E[Y q(Z,A,X)\mathbb{I}\{d_w^*(X,W) = A\} - h(W, \hat{d}_z(X,Z),X)|X=x] f(x)dx$$ 
is bounded by $O(n^{-\eta})$ and $\mathbb{P}(\mathcal{X}_{g1,2})$ converges to 0 in probability based on Assumption \ref{aszw} and the definition of $\mathcal{X}_{g1}$. A similar proof can also be conducted to obtain $\int_{x \in \mathcal{X}_{g2}} E[h(W, d_z^*(X,Z),X) - Y q(Z,A,X)\mathbb{I}\{\hat{d}_w(X,W) = A\}|X = x] f(x)dx$ is small enough.
We then have that $V(\hat{d}_{zw}^{\bar{\pi}}) \xrightarrow{p} V(d_{zw}^{\bar{\pi}*})$.

As we have proved that  $\mathbb{K}(\hat{\pi}) = V(\hat{d}_{zw}^{\bar{\pi}}) 
 - V(\hat{d}_{zw}^{\hat{\pi}}) = o(1)$ almost surely in Appendix \ref{sec:asy}, we finally conclude that $V(\hat{d}_{zw}^{\hat{\pi}}) \xrightarrow{p} V(d_{zw}^{\bar{\pi}*})$.

\section{Data generating mechanisim and parameter setup in Section~\ref{section5}}\label{sec:dgp}
The data generating mechanism for $(X,A,Z,W,U)$ is summarized in Table \ref{genetab}, and the setups of varying parameters in each scenario are summarized in Table~\ref{tab1}. 

\begin{sidewaystable}
   \centering 
   \small
   \begin{threeparttable}
     \begin{tabular}{ccc}
     Variables  & Generating Mechanism & Fixed Parameter Setting\\
      \midrule\midrule
 $X \in \mathbb{R}^2$ &   
 $X \sim \mathcal{N}(\Gamma_x, \Sigma_x)$    &   $\Gamma_x = (0.25, 0.25)^T, \Sigma_x = \left(\begin{array}{clr}
  0.25^2 & 0 \\
  0 & 0.25^2 \\	
  \end{array}\right)$                      \\
     \cmidrule(l  r ){1-3}
     $ A \in \{1,-1\}$ & $\left(\frac{A+1}{2}\right)|X \sim \text{Bern}(p_a)$ & $p_a =  \frac{1}{1+\exp\{(0.125,0.125)^TX\}}$ \\ 
     \cmidrule(l r ){1-3}
      $Z \in \mathbb{R}$ & 
  &  $\alpha_0 = \alpha_a = \mu_0 =  \kappa_0 =  \kappa_a = \sigma_{zw} = 0.25, $ \\ 
 $W \in \mathbb{R}$
     &   $(Z, W, U)|A, X \sim \mathcal{N}\left\{ \left ( \begin{array}{clr}
 \alpha_0+\alpha_a \frac{1+A}{2}+\alpha_x X\\
 \mu_0+\mu_a \frac{1+A}{2}+\mu_x X\\
 \kappa_0+\kappa_a \frac{1+A}{2}+\kappa_x X\\
 \end{array}\right), \Sigma = \left( \begin{array}{clr}
 \sigma^2_z & \sigma_{zw} & \sigma_{zu}\\
 \sigma_{zw} & \sigma_{w}^2 & \sigma_{wu}\\
 \sigma_{zu} & \sigma_{wu} & \sigma_{u}^2
 \end{array}
 \right)\right\}$
     &   $\mu_a = 0.125, \alpha_x = \mu_x  = \kappa_x = (0.25,0.25)^T,$\\
     $U \in \mathbb{R}$
     & 
  &   $ \sigma_{zu} = \sigma_{wu} = 0.5,  \sigma_z = \sigma_w = \sigma_u = 1$            \\
  \cmidrule(l r ){1-3}
  $Y \in \mathbb{R}$ & $Y \sim \mathcal{N}\{\mathbb{E}(Y | W, U, A, Z, X), \sigma_y^2\}$ & $\sigma_y = 0.25, b_0 = 2, \omega = 2$ \\
     \midrule\midrule
     \end{tabular}
     \begin{tablenotes}
 \item[*] As for generation of $Y$, $\mathbb{E}(Y|X,A,Z,W,U) = b_0 + b_{1}(X)\frac{1+A}{2} + b_2(X)X + \left(b_w +  b_a\frac{1+A}{2} + b_3(X)A -  \omega\right)\mathbb{E}(W|U,X) + \omega W$,
 where
 $\mathbb{E}(W|U,X) = \mu_0 + \mu_x X + \frac{\sigma_{wu}}{\sigma_u^2} (U - \kappa_0 - \kappa_x X).$
 \end{tablenotes}
 \end{threeparttable}
 \caption{Data generating mechanism and setup for fixed parameters across scenarios.}
 \label{genetab}
\end{sidewaystable}
 \begin{table*}[h]
     \centering
     \begin{tabular}{c|c|c|c|c|c|}
         \toprule
         \textbf{Scenario} & \multicolumn{5}{c|}{\textbf{Parameter Setup}} \\
         \cmidrule{2-6}
          \textbf{Number} &   $b_1(X)$  & $b_2(X)$ &$ b_3(X)$ & $b_a$ & $b_w$  \\
         \midrule
         \textbf{1} & $0.5 + 3X_{(1)} - 5X_{(2)}$ &  $(0.25, 0.25)^T$ & 0 & 0.25  & 8 \\
         \textbf{2} & $0.5 + 3X_{(1)} - 5X_{(2)}$ & $(0.25, 0.25)^T$  & 0 & 0 & 8 \\
         \textbf{3} &  $2.3 + |X_{(1)}-1| -|X_{(2)}+1|$ & $X^T$ & $\sin(X_{(1)}) -2\cos(X_{(2)})$ & -2.5 & 4 \\
         \textbf{4} &  $0.25 -6X_{(1)}X_{(2)}$ & $X^T$ & 0 & 0  & 5 \\
         \textbf{5} &  $0.1-2X_{(1)}^2$ & $X^T$ & $4X_{(2)}^2$ & 0.8  & 8 \\
         \textbf{6} &  $-0.5 + \exp(X_{(1)}) -3X_{(2)}$ & $(0.25, 0.25)^T$ & 0 & 0  & 8 \\
        \bottomrule
     \end{tabular}
     \begin{tablenotes}
 \item[*] *$X_{(1)}, X_{(2)}$ denote the first and second dimensions of $X$.
 \item[*] * The parameter settings in scenarios 1-4 are considered by \citet{qi2021proximal}.
 \end{tablenotes}
     \caption{The varying parameters for each scenario.}
     \label{tab1}
 \end{table*}

\section{Derivation of optimal ITRs considered in Section~\ref{section5}}\label{sec:dstar}
From 
$$(Z, W, U)|A, X \sim \mathcal{N}\left\{ \left ( \begin{array}{clr}
\alpha_0+\alpha_a \frac{1+A}{2}+\alpha_x X\\
\mu_0+\mu_a \frac{1+A}{2}+\mu_x X\\
\kappa_0+\kappa_a \frac{1+A}{2}+\kappa_x X\\
\end{array}\right), \Sigma = \left( \begin{array}{clr}
\sigma^2_z & \sigma_{zw} & \sigma_{zu}\\
\sigma_{zw} & \sigma_{w}^2 & \sigma_{wu}\\
\sigma_{zu} & \sigma_{wu} & \sigma_{u}^2
\end{array}
\right)\right\},$$ and
$$(Z, W, U)|X \sim \mathcal{N}\left\{ \left ( \begin{array}{clr}
\alpha_0+\alpha_a \mathbb{P}(A=1|X)+\alpha_x X\\
\mu_0+\mu_a \mathbb{P}(A=1|X)+\mu_x X\\
\kappa_0+\kappa_a \mathbb{P}(A=1|X)+\kappa_x X\\
\end{array}\right), \Sigma = \left( \begin{array}{clr}
\sigma^2_z & \sigma_{zw} & \sigma_{zu}\\
\sigma_{zw} & \sigma_{w}^2 & \sigma_{wu}\\
\sigma_{zu} & \sigma_{wu} & \sigma_{u}^2
\end{array}
\right)\right\},$$
the following results hold,
\begin{align}
	\mathbb{E}[W|X,A,U] &= \mu_0 + \mu_a \frac{1+A}{2} +\mu_x X +\frac{\sigma_{wu}}{\sigma_u^2}(U - \kappa_0 -\kappa_a\frac{1+A}{2} - \kappa_x X), \label{sol} \\
	\mathbb{E}[U|X,Z] &= \kappa_0 + \kappa_a \mathbb{P}(A=1|X) + \kappa_x X + \frac{\sigma_{zu}}{\sigma_z^2} (Z - \alpha_0 - \alpha_a \mathbb{P}(A=1|X) - \alpha_x X), \label{sol1} \\
	\mathbb{E}[U|X,W] &= \kappa_0 + \kappa_a \mathbb{P}(A=1|X) + \kappa_x X + \frac{\sigma_{wu}}{\sigma_w^2} (W - \mu_0 - \mu_a \mathbb{P}(A=1|X) - \mu_x X). \label{sol2} 
\end{align}
Recall that
\begin{align*}
	\mathbb{E}(Y|X,A,Z,W,U) &= b_0 + b_{1}(X)\frac{1+A}{2} + b_2(X)X + \left(b_w +  b_a\frac{1+A}{2} + b_3(X)A - \omega\right)\\
	&\left(\mu_0 + \mu_x X + \frac{\sigma_{wu}}{\sigma_u^2} (U - \kappa_0 - \kappa_x X)\right) + \omega W,
\end{align*}
then we can find that
\begin{align}
	\mathbb{E}(Y|X,A,Z,U) &= b_0 + b_{1}(X)\frac{1+A}{2} + b_2(X)X + \left(b_w +  b_a\frac{1+A}{2} + b_3(X)A -  \omega\right) \nonumber \\
	&\left(\mu_0 + \mu_x X + \frac{\sigma_{wu}}{\sigma_u^2} (U - \kappa_0 - \kappa_x X)\right) + \omega \mathbb{E}[W|X,A,Z,U], \nonumber \\
	&= b_0 + b_{1}(X)\frac{1+A}{2} + b_2(X)X + \left(b_w +  b_a\frac{1+A}{2} + b_3(X)A -  \omega\right) \nonumber \\
	&\left(\mu_0 + \mu_x X + \frac{\sigma_{wu}}{\sigma_u^2} (U - \kappa_0 - \kappa_x X)\right) + \omega \mathbb{E}[W|X,A,U], \nonumber \\
	&= b_0 + b_{1}(X)\frac{1+A}{2} + b_2(X)X + \left(b_w +  b_a\frac{1+A}{2} + b_3(X)A -  \omega\right)\nonumber \\
	&\left(\mu_0 + \mu_x X + \frac{\sigma_{wu}}{\sigma_u^2} (U - \kappa_0 - \kappa_x X)\right) + \omega\left(\mu_0  +\mu_x X +\frac{\sigma_{wu}}{\sigma_u^2}(U - \kappa_0  - \kappa_x X)\right),\nonumber \\
	&= b_0 + b_{1}(X)\frac{1+A}{2} + b_2(X)X + \left(b_w +  b_a\frac{1+A}{2} + b_3(X)A \right) \nonumber \\
    &\left(\mu_0 + \mu_x X + \frac{\sigma_{wu}}{\sigma_u^2} (U - \kappa_0 - \kappa_x X)\right), \label{exre1}
\end{align}
where the first equality is duo to Assumption~\ref{std}, and the second equality is due to \eqref{sol}, and  
\begin{align}
	\mathbb{E}(Y|X,A,W,U) &= 	\mathbb{E}(Y|X,A,Z,W,U)  \nonumber \\
	&= b_0 + b_{1}(X)\frac{1+A}{2} + b_2(X)X + \left(b_w +  b_a\frac{1+A}{2} + b_3(X)A -  \omega\right) \nonumber \\
	&\left(\mu_0 + \mu_x X + \frac{\sigma_{wu}}{\sigma_u^2} (U - \kappa_0 - \kappa_x X)\right) + \omega W, \label{exre2}
\end{align}
where the first equality is due to Assumption \ref{std}.
Furthermore, note that
\begin{align*}
\mathbb{E}[h(W,1,X)|X,Z,U] &= \mathbb{E}[h(W,1,X)|X,U]\\
&= \mathbb{E}[Y|X,A=1,U]  \\
&= \mathbb{E}[Y|X,A=1,Z,U]  \\
&= b_0 + b_{1}(X) + b_2(X)X + \left(b_w +  b_a + b_3(X) \right)\left(\mu_0 + \mu_x X + \frac{\sigma_{wu}}{\sigma_u^2} (U - \kappa_0 - \kappa_x X)\right),
\end{align*}
where the first and third equality is due to Assumption~\ref{std}, the second equality follows from Theorem~1 of \cite{miao2018identifying} under Assumptions~\ref{com1} and \ref{eofob}, and the last equality is by \eqref{exre1}.
Similarly,
\begin{align*}
\mathbb{E}[h(W,-1,X)|X,Z,U] &= \mathbb{E}[Y|X,A=-1,Z,U]  \\
&= b_0  + b_2(X)X + \left(b_w - b_3(X) \right)\left(\mu_0 + \mu_x X + \frac{\sigma_{wu}}{\sigma_u^2} (U - \kappa_0 - \kappa_x X)\right).
\end{align*}
On the other hand, 
\begin{align*}
\mathbb{E}[Yq(Z,1,X)\mathbb{I}\{A=1\}|X,W,U] &= \mathbb{P}(A=1|X,W,U)\mathbb{E}[Yq(Z,1,X)|X,A=1,W, U] \nonumber \\
&= \mathbb{P}(A=1|X,U)\mathbb{E}[q(Z,1,X)|X,A=1, U]\mathbb{E}[Y|X,A=1,W,U]\\
&= \mathbb{E}[Y|X,A=1,W,U] \\
&= b_0 + b_{1}(X) + b_2(X)X + \left(b_w +  b_a + b_3(X) -  \omega\right)\\
	&\left(\mu_0 + \mu_x X + \frac{\sigma_{wu}}{\sigma_u^2} (U - \kappa_0 - \kappa_x X)\right) + \omega W,
\end{align*}
where the second equality is due to Assumption~\ref{std}, and the third equality is due to Theorem 2.2 of \cite{cui2020semiparametric}  under Assumptions~\ref{com2} and \ref{eoftb}, and the last equality is due to \eqref{exre2}.
Similarly,
\begin{align*}
\mathbb{E}[Yq(Z,-1,X)\mathbb{I}\{A=-1\}|X,W,U] &= \mathbb{E}[Y|X,A=-1,W,U] \\
&= b_0 + b_2(X)X + \left(b_w - b_3(X) -  \omega\right)\\
	&\left(\mu_0 + \mu_x X + \frac{\sigma_{wu}}{\sigma_u^2} (U - \kappa_0 - \kappa_x X)\right) + \omega W.
\end{align*}
Then we can find that
\begin{align*}
\mathbb{E}[h(W,1,X) - h(W,-1,X)|X,Z,U] 	&= b_{1}(X) + (b_a+2b_3(X)) \left(\mu_0 + \mu_x X + \frac{\sigma_{wu}}{\sigma_u^2} (U - \kappa_0 - \kappa_x X)\right), 
\end{align*}
\begin{align*}
&\mathbb{E}[Yq(Z,1,X)\mathbb{I}\{A=1\}-Yq(Z,-1,X)\mathbb{I}\{A=-1\}|X,W,U] \nonumber \\
&= b_{1}(X) + (b_a+2b_3(X)) \left(\mu_0 + \mu_x X + \frac{\sigma_{wu}}{\sigma_u^2} (U - \kappa_0 - \kappa_x X)\right). 
\end{align*}
Furthermore, we have
\begin{align}
\mathbb{E}[h(W,1,X) - h(W,-1,X)|X,Z] &=  \mathbb{E}[\mathbb{E}[h(W,1,X) - h(W,-1,X)|X,Z,U]] \nonumber \\
 	&= b_{1}(X) + (b_a+2b_3(X)) \left(\mu_0 + \mu_x X + \frac{\sigma_{wu}}{\sigma_u^2} (\mathbb{E}[U|X,Z] - \kappa_0 - \kappa_x X)\right), \label{re3}
\end{align}
\begin{align}
&\ \ \ \ \mathbb{E}[Yq(Z,1,X)\mathbb{I}\{A=1\}-Yq(Z,-1,X)\mathbb{I}\{A=-1\}|X,W] \nonumber \\
&=\mathbb{E}[\mathbb{E}[Yq(Z,1,X)\mathbb{I}\{A=1\}-Yq(Z,-1,X)\mathbb{I}\{A=-1\}|X,W,U]] \nonumber \\
&= b_{1}(X) + (b_a+2b_3(X)) \left(\mu_0 + \mu_x X + \frac{\sigma_{wu}}{\sigma_u^2} (\mathbb{E}[U|X,W] - \kappa_0 - \kappa_x X)\right). \label{re4}
\end{align}
Therefore, plug~\eqref{sol1} and \eqref{sol2} into \eqref{re3} and \eqref{re4} respectively, we can find that
\begin{align*}
\mathbb{E}[h(W,1,X) - h(W,-1,X)|X,Z] &= b_{1}(X) + (b_a+2b_3(X)) (\mu_0 + \mu_x X + \frac{\sigma_{wu}}{\sigma_u^2} (\kappa_0 + \kappa_a \mathbb{P}(A=1|X) \\
&+ \kappa_x X + \frac{\sigma_{zu}}{\sigma_z^2} (Z - \alpha_0 - \alpha_a \mathbb{P}(A=1|X) - \alpha_x X) - \kappa_0 - \kappa_x X)), 
\end{align*}
\begin{align*}
&\ \ \ \ \mathbb{E}[Yq(Z,1,X)\mathbb{I}\{A=1\}-Yq(Z,-1,X)\mathbb{I}\{A=-1\}|X,W] \\
&= b_{1}(X) + (b_a+2b_3(X)) (\mu_0 + \mu_x X + \frac{\sigma_{wu}}{\sigma_u^2} (\kappa_0 + \kappa_a \mathbb{P}(A=1|X) \\
&+ \kappa_x X + \frac{\sigma_{wu}}{\sigma_w^2} (W - \mu_0 - \mu_a \mathbb{P}(A=1|X) - \mu_x X) - \kappa_0 - \kappa_x X)).
\end{align*}
Hence, 
\begin{align*}
	d_z^*(X,Z) &= \text{sign}\{\mathbb{E}[h(W,1,X) - h(W,-1,X)|X,Z]\} \\
	&= \text{sign}\{b_{1}(X) + (b_a+2b_3(X)) (\mu_0 + \mu_x X + \frac{\sigma_{wu}}{\sigma_u^2} (\kappa_0 + \kappa_a \mathbb{P}(A=1|X) \\
&+ \kappa_x X + \frac{\sigma_{zu}}{\sigma_z^2} (Z - \alpha_0 - \alpha_a \mathbb{P}(A=1|X) - \alpha_x X) - \kappa_0 - \kappa_x X)) \}, \\
	d_w^*(X,W)  &= \text{sign}\{\mathbb{E}[Yq(Z,1,X)\mathbb{I}\{A=1\}-Yq(Z,-1,X)\mathbb{I}\{A=-1\}|X,W]\} \\
	&= \text{sign}\{b_{1}(X) + (b_a+2b_3(X)) (\mu_0 + \mu_x X + \frac{\sigma_{wu}}{\sigma_u^2} (\kappa_0 + \kappa_a \mathbb{P}(A=1|X) \\
&+ \kappa_x X + \frac{\sigma_{wu}}{\sigma_w^2} (W - \mu_0 - \mu_a \mathbb{P}(A=1|X) - \mu_x X) - \kappa_0 - \kappa_x X))\}.
\end{align*}

\section{Implementation details of numerical experiments}
\label{sec:im}
\textbf{Step (i)}
The method we adopt is neural maximum moment restriction (NMMR), which employs multilayer perceptron (MLP) to estimate the confounding bridges \citep{kompa2022deep}. The target loss functions are set as
$$R(h) = \mathbb{E}[(Y-h(W,A,X))(Y' - h(W',A',X'))K_z((Z,A,X),(Z',A',X'))],$$
$$R(q, a) = \mathbb{E}[(1-\mathbb{I}\{A = a\}q(Z,a,X))(1-\mathbb{I}\{A' = a\}q(Z', a, X'))K_w((W, X), (W', X'))], \text{~for} \ a \in \mathcal{A},$$
where $(Z', W', A', X', Y')$ are independent copies of $(Z,W,A,X,Y)$, and $K_z:(\mathcal{Z} \times \mathcal{A} \times \mathcal{X})^2 \rightarrow \mathbb{R}, K_w:(\mathcal{W} \times \mathcal{X})^2 \rightarrow \mathbb{R}$ denote continuous, bounded, and integrally strictly positive definite (ISPD) kernels. In practice, we use the empirical risk  instead, i.e.,
\begin{align}
\hat{R}(h) &= \frac{1}{n(n-1)}\sum_{i,j=1, i \neq j}^n (y_i - h_i)(y_j - h_j)k_{z,ij}, \label{lossh}\\
\hat{R}(q, a) &=\frac{1}{n(n-1)}\sum_{i,j=1, i \neq j}^n (1-\mathbb{I}\{a_i = a\}q_i)(1-\mathbb{I}\{a_j = a\}q_j)k_{w,ij}, \text{~for} \ a \in \mathcal{A}, \label{lossq}
\end{align}
where $h_i =h(w_i, a_i, x_i), q_i = (z_i, a_i, x_i), k_{z,ij} = K_z((z_i, a_i, x_i)(z_j, a_j, x_j))$ and $k_{w, ij} = K_w((w_i, x_i),(w_j, x_j))$. 
In addition, we add a penalty term with respect to network weights to avoid overfitting.

As for the hyperparameters tuning procedure, we consider employing multilayer perceptrons with 2-8 fully connected layers with a variable number of hidden units. We then perform a grid search over the following parameters: learning rate, penalty coefficient, number of epochs, batch size, depth of the network, and width of the network. For every permutation of these parameters, we train a network based on the determined architecture and parameter values. Subsequently, we compute the empirical risk. Our aim is to pinpoint the parameter combination that yields the lowest empirical risk. These identified optimal parameters are then utilized to construct a refined neural network, which, in turn, serves as the foundation for conducting estimations. The parameter setup is summarized in Table~\ref{NNpara}. For detailed insights into the specific hyperparameter choices and architectural dimensions, we refer to supplementary Section B in \cite{kompa2022deep}.
\begin{table}[htbp]
\centering
\begin{tabular}{|c|l|}
\hline
    \textbf{Parameter} &  \textbf{Value} \\
    \hline
	Number of epoch & 150\\
	Batch size & 250\\
	Learning rate & 0.003\\
	Penalty coefficient&  0.001, 0.01, 0.1 \\
	Depth of network & 4 (for estimating $h$) \\
     & 8 (for estimating $q$)\\
    Width of network & 80\\
    \hline
\end{tabular}
\caption{Parameter setup for step (i)}
\label{NNpara}
\end{table}

\textbf{Step (ii)} For the estimation of preliminary ITRs, we follow the main text to solve the proposed optimization problems. For instance, to estimate $d_z^*$, we solve the following optimization problem:
$$\hat{g}_z \in \arg\min_{g_z \in \mathcal{G}_z} \mathbb{P}_n[\{\hat{h}(W,1,X)-\hat{h}(W,-1,X)\} \phi(g_z(X,Z))] + \rho_{z}||g_z||_{\mathcal{G}_\mathcal{Z}}^2.$$
Here, $g_z$ represents a measurable decision function in $\mathcal{G}_\mathcal{Z} : \mathcal{X} \times \mathcal{Z} \rightarrow \mathbb{R}$ used to indicate $d_z$ (e.g., $d_z(X,Z) = \text{sign}(g_z(X,Z))$), $\phi$ denotes the hinge loss function $\phi(x) = \max\{1 - x, 0\}$, and $\rho_z > 0$ is a tuning parameter. As for the tuning procedure regarding $\rho_z$, when $g_z$ is treated as a linear rule, for each predefined $\rho_z$, the data is divided into $K$ folds. For each $k \in [K]$, we compute $\hat{h}^{(-k)}$ and $\hat{g}_z^{(-k)}$, and then calculate the empirical value using the validation data. By averaging the empirical values across $K$ folds for each value of $\rho_z$, we identify the parameter that maximizes the average empirical value. The finalized parameter is then employed to determine $\hat{g}_z$. Such a procedure can be extended. For example, when considering $g_z$ as a RKHS, it is advisable to apply the cross-fitting procedure separately for each combination of pre-defined $\rho_z$ and bandwidth, with details presented in \cite{qi2021proximal}. And the estimation of $\hat{d}_w$ can be approached in a similar manner.

For more estimators regarding $d_z^*$ and $d_w^*$, we refer to \cite{bennett2021proximal,  sverdrup2023proximal, wang2022blessing}. One could further expand the estimation pipeline utilized in unconfounded scenarios and leverage state-of-the-art machine learning techniques \citep{chen2020representation, raghu2017continuous, yoon2018ganite} to tackle the weighted classification problems and construct estimates.

\textbf{Step (iii)} The estimation of $\bar{\pi}$ follows the procedure given in the main text. As for the selection of bandwidth in the Nadaraya-Watson kernel regression estimator, we employ Scott’s rule of thumb \citep{scott2015multivariate} and set $\gamma = 1.06 \hat{\sigma} n^{-1/5}$, where $\hat{\sigma}$ is the estimated standard deviation of $X$. For more methods regarding estimation of $\delta(\cdot)$, we refer to \cite{chen2017tutorial, dalmasso2020conditional, dinh2016density, sohn2015learning}.

For the convenience of readers to reproduce the results, the pseudo-code of the whole pipeline is presented in Algorithm \ref{algo}. The code of implementation can also be accessed on GitHub~\footnote{\scriptsize \url{https://github.com/taoshen2022/Optimal-Treatment-Regimes-for-Proximal-Causal-Learning}}.

\begin{algorithm}[htbp]
\caption{Estimation of optimal ITR $d_{zw}^{\bar{\pi}*}$}
\label{algo}
\textbf{Input}: Training data \\
Construct MLP models to estimate $h(w,a,x)$ and $q(z,a,x)$:\\
\textbf{Repeat} for different penalty coefficients:\\
\ \ \ \ \For{\text{each epoch}}{
\ \ \ \ \For{\text{each batch}}{
\ \ \ \  Compute loss function \eqref{lossh} and \eqref{lossq} based on the batch\\
\ \ \ \  Update the internal model parameter
}
}
\textbf{Finalize} the penalty coefficient which minimizes the empirical loss, and obtain $\hat{h}(w,a,x)$ and $\hat{q}(z,a,x)$\\
\textbf{Repeat} for different $\rho_{z}$ and $\rho_{w}$:\\
\ \ \ \ \For{\text{each batch}}{
\ \ \ \  Find $\hat{g}_{z,b}, \hat{g}_{w,b}$ by \eqref{eq:minz} and \eqref{eq:minw} based on the $b$-th batch, estimated bridge functions, and specified $\rho_{z}$ and $\rho_{w}$, and then obtain $\hat{d}_{z,b}, \hat{d}_{w,b}$ based on $\hat{g}_{z,b}, \hat{g}_{w,b}$\\
\ \ \ \ Compute empirical value of $\hat{d}_{z,b}, \hat{d}_{w,b}$  respectively using the data not covered in the batch
}
\textbf{Finalize} $\rho_{z}$ and $\rho_{w}$ based on empirical values and then obtain $\hat{d}_z, \hat{d}_w$\\
Select bandwidth by Scott's rule of thumb\\
Find $\hat{\delta}(X;\hat{d}_z,\hat{d}_w)$ and then obtain $\hat{\pi}(X;\hat{d}_z,\hat{d}_w)$\\
\textbf{Output}: $\hat{d}_{zw}^{\hat{\pi}}$ constructed by \eqref{eq:est}
\end{algorithm}

\section{Additional results of numerical experiments}
\begin{figure*}[htbp]
        \centering
        \includegraphics[width=0.9\textwidth, height = 0.5\textwidth]{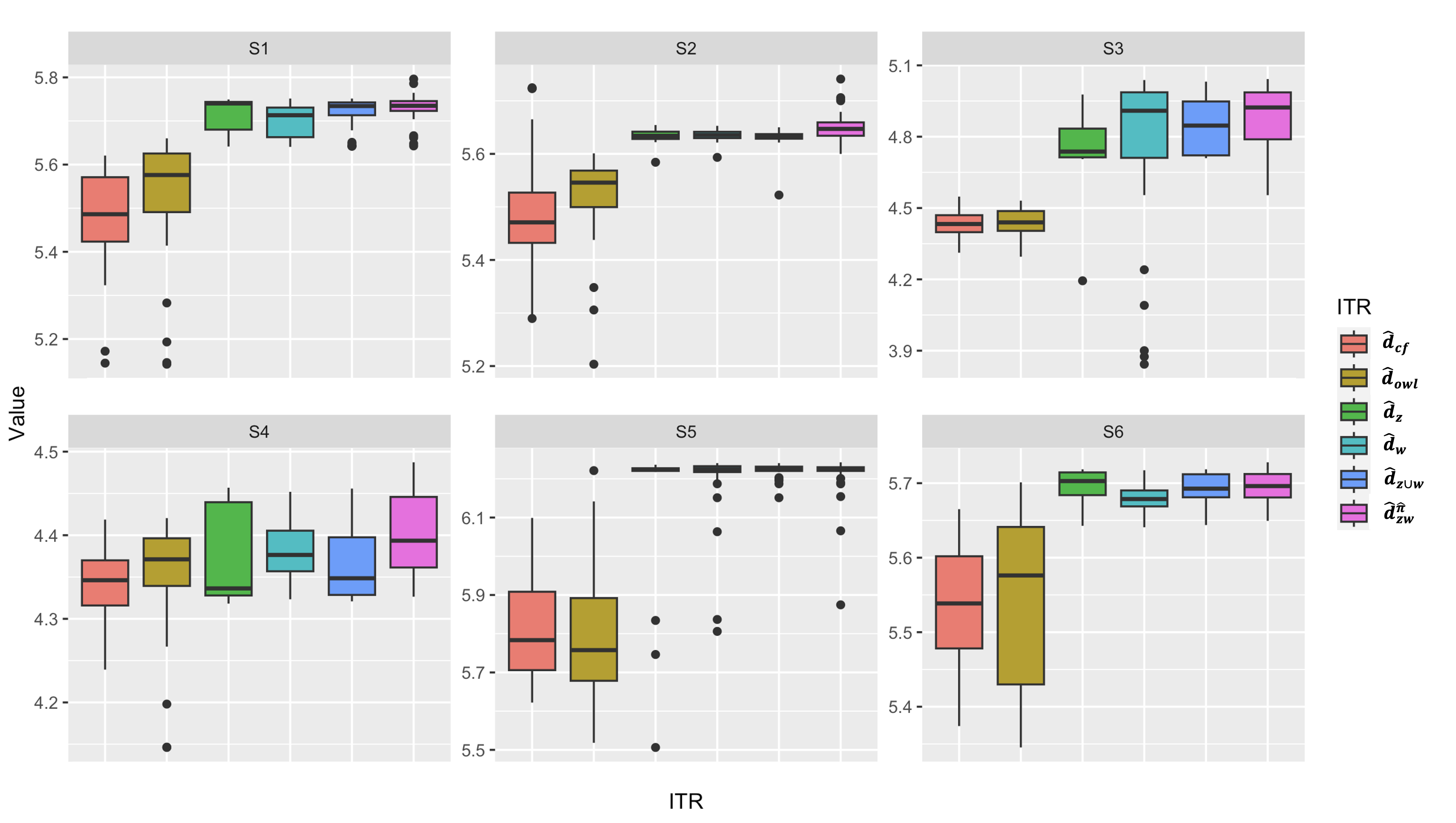}
        \caption{Boxplots of the empirical value functions with $n=500$.}
        \label{figad1}
\end{figure*}

\begin{figure*}[htbp]
        \centering
        \includegraphics[width=0.9\textwidth, height = 0.5\textwidth]{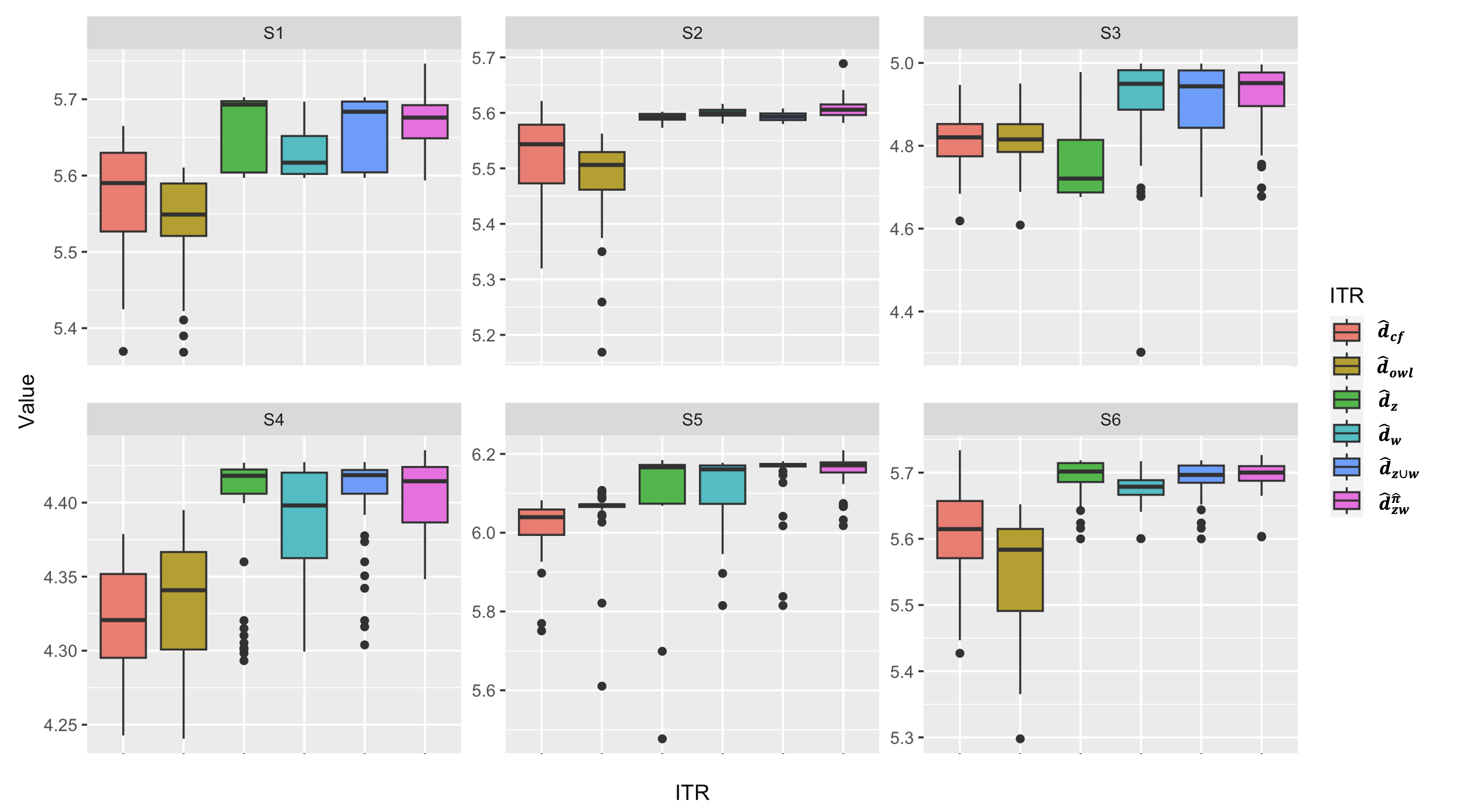}
        \caption{Boxplots of the empirical value functions with $n=500$ and an altered behavior policy.}
        \label{figad2}
\end{figure*}

The experimental results with sample size $n=500$ are presented in Figure~\ref{figad1}.
The experimental results with sample size $n=500$ and an altered behavior policy (treatment is randomly assigned in this case) are presented in Figure~\ref{figad2}. 
\label{sec:add}

\section{Additional results of real data application}
\label{sec:add2}
Regarding the quantitative analysis, Table~\ref{tabreal} describes the estimated value functions of our proposed ITR, alongside existing approaches, under four settings with increasing numbers of proxies. For Setting~1, $Z = (pafi1, paco21)$, $W = (ph1, hema1)$. For Setting~2, $Z = (pafi1,paco21,pot1)$, $W = (ph1,hema1,bili1)$. For Setting~3, $Z = (pafi1,paco21,pot1,wt0)$, $ W= (ph1,hema1,bili1,sod1)$. For Setting~4, $Z = (pafi1,paco21,pot1,wt0,crea1)$, $W= (ph1,hema1,bili1,sod1,alb1)$.

\begin{table}[htbp]
	\centering
 \begin{threeparttable}
	\begin{tabular}{lp{1.6cm}p{1.6cm}p{1.6cm}p{1.6cm}p{1.6cm}p{1.6cm}}
		\toprule  
		& $\hat{V}(\hat{d}_{cf})$ & $\hat{V}(\hat{d}_{owl})$ & $\hat{V}(\hat{d}_z)$ & $\hat{V}(\hat{d}_w)$ & $\hat{V}(\hat{d}_{z \cup w})$ & $\hat{V}(\hat{d}_{zw}^{\hat{\pi}})$  \\
		\midrule  
		Setting 1 & 24.84 (3.06) & 24.97 (2.93) & 25.12 (4.69) & 26.61 (3.34) & 27.86 (2.28) & 28.21 (3.28)  \\
	  Setting 2 &  24.81 (3.11) & 24.97 (2.94) & 25.60 (3.73) & 25.74 (3.57) & 26.32 (2.29) & 27.02 (2.95) \\
        Setting 3 & 24.79 (3.02) & 24.97 (2.93) & 26.12 (3.61) & 25.53 (3.29) & 26.76 (2.76) & 27.83 (3.03)  \\
        Setting 4 &  24.90 (3.18) & 24.97 (2.93)& 25.26 (4.76) & 25.81 (3.03) & 27.38 (2.74) & 27.96 (3.07)\\
	\bottomrule
	\end{tabular}
 \end{threeparttable}
 \caption{Estimated values for different ITRs under different proxy variable settings.}
\label{tabreal}
\end{table}

As for the qualitative analysis, we present an illustrative example below. Regarding the estimated ITRs in Setting 1, the coefficient of $cat1\_lung$ is negative with a minor magnitude for $\hat{d}_z$, contrasting with a positive and relatively large coefficient observed for $\hat{d}_w$, which mirror the outcomes outlined in \cite{qi2021proximal}. This finding suggests that, within the primary disease category of patients with lung cancer, $\hat{d}_z$ advocates for undergoing RHC, while $\hat{d}_w$ displays a notably inconclusive trend. As evidenced by 
$\hat{\pi}$, the prevailing trajectory for patients with $cat1\_lung = 1$ involves a strong inclination toward undergoing RHC, i.e., $\hat{\pi}(X) = 1$, aligning with the guidance offered by $\hat{d}_z$. Significantly, the domain knowledge underscores the potential for patients with advanced lung cancer to develop complications like pulmonary hypertension and coma, potentially warranting RHC for assessing pulmonary vascular changes and informing treatment strategies \citep{galie2009guidelines}, which lends support to the recommendations offered by our proposed regime. Furthermore, it is important to note that the whole group of patients can be regarded as unions of multiple subgroups based on various distinct features, and the superiority of $\hat{d}_w$ is evident in some subgroups (e.g., $amihx$). These results show that our proposed ITR offers superior efficacy compared to $\hat{d}_z, \hat{d}_w$ and $\hat{d}_{z\cup w}$ as our methodology incorporates selection through $\hat{\pi}$.


\end{document}